\documentclass[a4paper,11pt,reqno,superscriptaddress,nofootinbib, showkeys, aps, pre]{revtex4-2}
\usepackage[centertags]{amsmath}
\usepackage{amsfonts}
\usepackage{amssymb}
\usepackage{amsthm}
\usepackage{titlesec}
\usepackage{newlfont}
\usepackage{stmaryrd}
\usepackage{mathrsfs}
\usepackage{mathtools}
\usepackage{euscript}
\usepackage{graphicx}
\usepackage{enumerate}
\usepackage{todonotes}
\usepackage{color}
\usepackage{orcidlink}
\usepackage{subcaption}
\usepackage{booktabs}
\usetikzlibrary{decorations.pathmorphing, arrows.meta}
\usepackage{tikz}
\usepackage{pgf}
\usetikzlibrary{positioning,fit,calc}
\usetikzlibrary{arrows,automata}
\usepackage{wrapfig}
\usepackage{amscd}
\usepackage{cancel}
\usepackage{hhline}
\usepackage{import}
\usepackage{enumitem} 
\newlist{condenum}{enumerate}{1} 
\setlist[condenum]{label=\bfseries Condition \arabic*., 
          ref=\arabic*, wide}

\usepackage{changes}
\date{\today}
\usepackage{float}

\let\oldsqrt\sqrt
\def\sqrt{\mathpalette\DHLhksqrt}
\def\DHLhksqrt#1#2{
	\setbox0=\hbox{$#1\oldsqrt{#2\,}$}\dimen0=\ht0
	\advance\dimen0-0.2\ht0
	\setbox2=\hbox{\vrule height\ht0 depth -\dimen0}
	{\box0\lower0.4pt\box2}}

\newcommand{\id}{\textrm{d}}

\usepackage{url}
\usepackage{lipsum}
\usepackage{mathtools}
\usepackage{lmodern}
\usepackage{anyfontsize}
\usepackage{hyperref}

\begin{document}
\title{A Rayleigh criterion for mechanical instability: 
inducing~activity~by~chemo–mechanical~coupling}

\author{Aaron Beyen \orcidlink{0000-0002-4341-7661}}
\affiliation{Department of Physics and Astronomy, KU Leuven}

\author{Francesco Casini \orcidlink{0000-0001-5245-7297}}
\affiliation{Department of Physics and Astronomy, KU Leuven}

\author{Christian Maes \orcidlink{0000-0002-0188-697X}}
\affiliation{Department of Physics and Astronomy, KU Leuven}

\begin{abstract}
\noindent Instabilities in thermodynamic systems are often undesirable, as they can lead to loss of control or even catastrophic behavior. Yet, the same mechanisms can also generate rich nonequilibrium behavior and may play a constructive role in living systems. We introduce a theoretical framework, inspired by Rayleigh’s analysis of thermoacoustic instabilities, to study the emergence of mechanical activity. In particular, we derive Rayleigh-like criteria governing the onset of activity and the generation of rotational motion in a slow Newtonian probe coupled to driven chemical processes, described by Markov jump processes. These criteria are expressed in terms of the phase relation between entropic and frenetic contributions, providing a transparent condition for when chemical driving results in sustained rotational or active mechanical motion.
\end{abstract}


\maketitle

\section{Introduction}
In classical physics, instabilities often mark the onset of motion. A paradigmatic example is the thermoacoustic instability analyzed by Rayleigh, where heat input, when appropriately phased with pressure oscillations, drives self-sustained sound, \cite{Rayleigh1878Explanation, Sujith2021Thermoacoustic}. This mechanism illustrates a general principle: energy supplied out of equilibrium can generate coherent mechanical motion when the coupling obeys specific phase relations \cite{Rayleigh01041883}.\\
A closely related question arises in biophysical systems. There, mechanical activity is not imposed externally but emerges from underlying chemical processes that continuously dissipate energy. Molecules consume fuel, reactions proceed out of equilibrium, and yet the conditions under which this chemical driving produces persistentmechanical motion remain only partially understood. What plays the role of Rayleigh’s phase criterion in such chemo-mechanical systems?\\

In this work, we address that question at a general level. Complementary to modeling more coarse-grained biological and hydrodynamical details of chemically driven active systems, as in \cite{janusparticles, diffusionphoreticswimmer, stochasticmicroswimmer}, we develop a bottom-up thermodynamically consistent framework that captures how nonequilibrium chemical driving couples to mechanical degrees of freedom, following examples that include recent work on active solids \cite{Cocconi_2025}, flocking systems \cite{Agranov_2024,Agranov_2025}, and the modelling of active particles \cite{Fritz_2023}. What is new is that we combine local detailed balance with semi-reciprocity to derive Rayleigh-like criteria for the emergence of activity, expressed in terms of the phase relation between entropic and frenetic contributions. This provides a physically transparent condition for when driven chemistry generates sustained mechanical motion, the basic example of bio-chemo-mechanics, \cite{Sun2022BiochemoMech,Feng2025MechanoChemoBio}.\\
Semi-reciprocity guides the feedback between two sets $(x,\sigma)$ of degrees of freedom, possibly of a very different nature. It refers to the situation where the mutual forces that actually couple the two systems originate from the same interaction energy $U(x,\sigma)$, but not necessarily of the form $U(x,\sigma) = u(x-\sigma)$ (full reciprocity) since $x$ and $\sigma$ might live in different spaces.\\ 
Local detailed balance, on the other hand, prescribes how to enter the thermodynamic forces that bias microscopic transitions, and how entropy and energy changes in the environment (at uniform temperature) enter the system trajectories, \cite{ldb, ldb3,KatzLebowitzSpohn1984,hal}. Fundamentally, credible models of chemo–mechanical activity must respect those two principles; see \cite{howfaractive,seifert1, seifert2, julicher1997modeling, Basu_2015,nonreciprocalmanybodyphysics} for issues and discussions. Otherwise, one may transfer or generate rotational motion, even activity, but not in a way that can be cleanly and consistently interpreted mechanically and thermodynamically. 
Reconstructing these ways is part of ongoing and major research initiatives in experimental and theoretical mechanobiology, also related to pattern formation and mechanically induced biological functioning, 
\cite{vrugt2025exactlyactivematter,Gompper2020,activeparticle1,activematter,Seifert2011,Wurthner2022BridgingScales,Burkart2022ControlProteinPatterns}.\\

To keep the discussion sharp and simple, we work in the deliberately restricted setup of a probe (point-particle) moving on a circle, coupled to a collection of fast, independent but driven jump processes, abbreviated as ``jumpers''. Similar switching-state dynamics have long been used to model flashing and rocking ratchets, molecular motors, chemically driven active particles, and diffusion in randomly switching energy landscapes \cite{seif, modellingmolecularmotors, mmotors, Molecularcombustionmotors, Parrondo1998ReversibleRatchets, ParrondoDinis2007RatchetsParadoxicalGames, FluxReversalinaSymmetricOpticalThermalRatchet, dichtratch, forcedthermalratch, Diffusionrandompotential, Diffusionrandompotential2}, where the switching corresponds to transitions between distinct energy landscapes. In contrast, the focus here is not transport or rectification, but directed at the transfer of nonequilibrium features and the mechanism of activity generation in particular. The jumpers act as an internal chemical reservoir, driven away from equilibrium, to which the probe is reciprocally coupled, all at the same finite temperature with the chemistry at finite driving affinity. Going beyond the result in \cite{MaesNetocny2019} where only the (leading) induced streaming term was discussed, we study here the emergence of friction and noise (at higher order) from such a coupling, and how the reduced probe dynamics becomes active, by a simple yet consistent (thermo-)dynamical coupling to driven chemistry. In these terms, our central result is that the chemo–mechanical coupling can produce a regime of (partially) negative linear friction over a range of parameters, \textit{i.e.} an effective anti-damping where the velocity amplitude grows; see also \cite{beyen2025couplingelasticstringactive, PeiMaes2025, pei2026transferActiveMotion} for a similar phenomenon in different settings. It is precisely the time-symmetric dynamical activity of the chemical section, often referred to as frenesy \cite{frenesy, fren1, Gaspard_2022, nondiss} and to be compared with the antisymmetric, dissipative entropy production part of the response, that allows chemical driving to feed energy into mechanical motion, at least under the appropriate phase condition.
That specification is new and original even though the phenomenon itself has been described at least since Rayleigh's work on the thermo-acoustic (linear) instability, \cite{Rayleigh1878Explanation,Sujith2021Thermoacoustic}, where \textit{e.g.} an amplification results if, on average, heat addition occurs in phase with the pressure increases during the oscillation. Nonlinearities typically cure this instability, and the complex transient and stationary behaviour can be represented by the simplified model of a Rayleigh oscillator, \cite{Rayleigh01041883,Chen_1994, quantumrayleigh}, in which negative linear friction at small velocities is balanced by nonlinear damping at larger amplitudes.
This opens the door to limit cycles generated purely by chemo–mechanical feedback, without externally imposed periodic forcing. \\

\vspace{-0.38 cm}

\underline{Plan of the paper}:  
We start with Section \ref{subsection model definition}, where the theoretical setup of a Newtonian probe dynamics coupled to Markov jump processes is introduced. Under the assumption of time-scale separation, we obtain and discuss in Section \ref{section reduced dynamics} the reduced dynamics of the probe when integrating out the bath degrees of freedom. The effective dynamics contains an induced mean force, a friction coefficient, and a noise term, which are computed explicitly under weak coupling. We discuss the Rayleigh-like criteria for the linear instability. In a way, the probe may start to behave as a Rayleigh oscillator, which stands for persistent mechanical behavior. The saturation and nonlinear regime (beyond time-scale separation) are analyzed in Section \ref{section saturation regime}. It includes the active and the rotational regimes, which are summarized in a number of numerical simulation results. The final Section \ref{section remarks} adds some additional remarks on the large driving limit, the emergence of nonreciprocal forces, and the possible generalization of a confined probe not moving on a circle. The Appendix contains various clarifications and explicit computations, together with a guide for the numerical implementation. 

\section{Model definition}\label{subsection model definition}

\begin{figure}[H]
  \centering
  \includegraphics[width=0.6\linewidth]{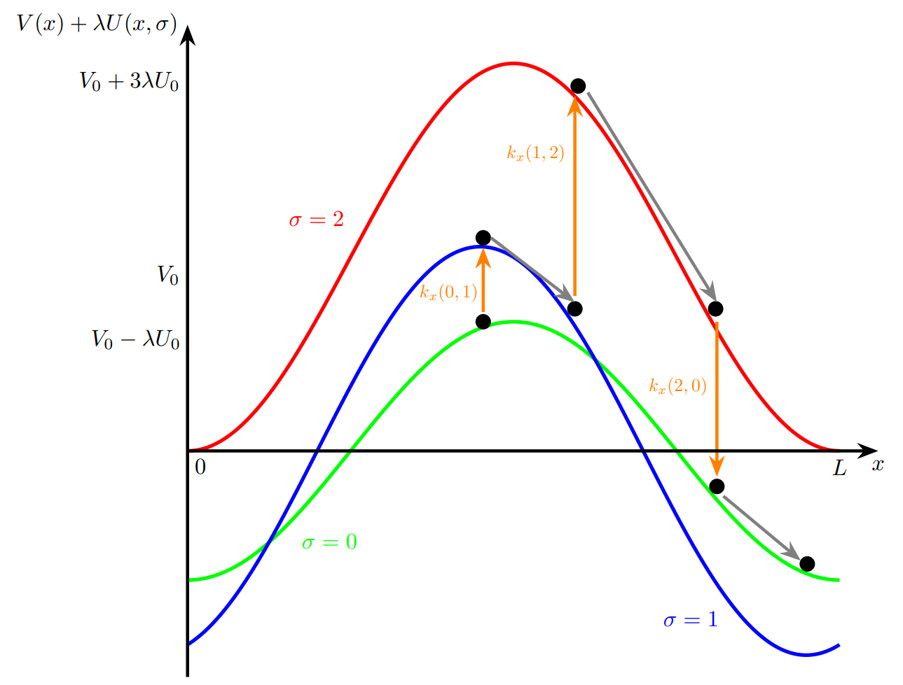}
   \caption{\raggedright Probe (black dot) following the dynamics \eqref{eq:probe_dyn} for $V(x) = -V_0 \cos\left(\frac{2 \pi}{L} x \right)$ and a 3-state switching potential $U(x, 0) = U_0 \cos\left(\frac{2 \pi }{L} x \right)$, $U(x,1) = U_0 \sin\left(\frac{2 \pi }{L} x \right), \ U(x,2) = 3 U_0$ with parameters $ \ \lambda = 0.5, V_0 = 3, U_0 = 2$. The grey arrows indicate the motion inside a potential, while the instantaneous jumps between potentials with rates $k_x(\sigma, \sigma')$ are represented by orange arrows. 
   }
  \label{fig switching potential}
\end{figure}

We consider a Newtonian (inertial) probe particle of mass $m$ moving on a circle of length $L$ with coordinate $x(t) \in [0, L)$ and velocity $v(t)$. The probe is coupled to a (chemical) bath consisting of $N$ independent copies of a driven
$n$–state Markov jump process (``jumpers'') where each jumper $i=1,\dots,N$ has internal state $\sigma_{i}(t) \in \mathbb{Z}_n = \{0,1,2,3, ...,n-1\}$. For the probe dynamics, we take
\begin{align}
\dot x(t) &= v(t), \qquad m \dot v(t) = - \frac{\partial V}{\partial x}(x(t)) 
+ \lambda \sum_{i=1}^N F\!\left(x(t),\sigma_{i}(t)\right)
\label{eq:probe_dyn}
\end{align}
with coupling constant $\lambda > 0$ and periodic self-potential $V(x) = -V_0 \cos \left( \frac{2 \pi}{L} x\right)$. 
The force $F$ encodes the
chemo–mechanical coupling, arising in a semi-reciprocal way $F(x,\sigma)= -\partial_x U(x,\sigma)$ with potential $U$, unspecified so far but periodic in $x$. Consequently, the mechanical particle evolves in the effective potential $V(x)+U(x,\sigma)$, whose form depends on the current chemical state $\sigma\in\mathbb Z_n$. The resulting impulsive dynamics, illustrated schematically in Fig.~\ref{fig switching potential}, is closely related to models of molecular motors, flashing and rocking ratchets, and diffusion in switching potentials mentioned before. Those models often focus on two-state switching ($n=2$), combined with spatially asymmetric potentials to generate directed transport. Here, by contrast, we consider at least three chemical states ($n\ge 3$), to break the detailed balance condition for the medium, and symmetric potentials, following ideas similar to those in \cite{FluxReversalinaSymmetricOpticalThermalRatchet, AsymmetricCycling}. \\
While many active and biological systems are naturally described in the overdamped (high friction) limit due to a low Reynolds number, our primary goal is to derive the reduced dynamics of a microscopic system (probe) in a medium. In that context, it is natural to start from the conservative (Newtonian) probe dynamics \eqref{eq:probe_dyn} and subsequently integrate out the nonequilibrium bath degrees of freedom. Indeed, that is also how a fundamental derivation of Brownian motion starts, often by applying the standard tools for deriving reduced dynamics, \textit{e.g.}, projection operators \cite{Mori1, Mori2, zwanzig, Schilling,widder2026generalised, grabert1982projection}, and singular perturbation theory \cite{singularperturbationtheory, singularescaperate, Tanogami2022, Tikhonov1952, Lomov1992}. At any rate, the overdamped limit is subtle and, in some situations, is best taken \emph{after} deriving the effective dynamics rather than at the level of the microscopic equations \cite{kamp, entropyproductionbrownianellipsoid, Sancho1982, Hottovy2012, Brownianinhomenv}. It does not mean that the larger (thermal) environment does not add friction and noise as well, but those are left unspecified here. An interesting future perspective is to compare our results with more standard Hamiltonian flashing ratchets, \cite{hamratchet1, hamratchet2}. \\
 
Coming back to \eqref{eq:probe_dyn}, for $\lambda = 0$, we have a typical velocity scale $v_c = \sqrt{ |V_0| /m}$ for the free motion, while a factorized potential $U(x, \sigma) = f(\sigma) \ U_0 \cos \left( \frac{2 \pi}{L} x\right)$ corresponds to a physical pendulum with effective gravitational force
\begin{align*}
  m\,g_{\text{eff}}(t) = \frac{2 \pi}{L} \left( V_0 - \lambda U_0 \sum_{i = 1}^N f(\sigma_i(t)) \right)
\end{align*}
Each jump process $\sigma_{i}(t)$ in \eqref{eq:probe_dyn} is independent and identically distributed for each $i$, and evolves following $\sigma \to \sigma'$ with transition rate $k_{x}(\sigma, \sigma')$
that depends parametrically on the instantaneous 
position $x$ of the probe via the same energy function $U(x,\sigma)$.
More precisely, to allow a separation of kinetic and thermodynamic input, we write the rates $k_x(\sigma, \sigma') = \psi_x(\sigma, \sigma') \ e^{s_x(\sigma, \sigma')/2}$ in terms of
\begin{align}
  \psi_x(\sigma, \sigma')= \psi_x(\sigma', \sigma) = \sqrt{k_x(\sigma, \sigma') \ k_x(\sigma', \sigma)},\qquad s_x(\sigma, \sigma')=-s_x(\sigma', \sigma) = \log \left( \frac{k_x(\sigma, \sigma')}{k_x(\sigma', \sigma)} \right) \label{rate decomposition}
\end{align}
with time-symmetric part $\psi_x$ (reactivities) and time-antisymmetric part $s_x$. It is worth emphasizing that the reactivities can depend on the probe position $x$, \cite{seifert1, julicher1997modeling}, and that will play an important role for the appearance of activity. 

With \eqref{rate decomposition} in mind, following \cite{frenetic_Steering}, we take 
\begin{align}\label{driv}
  s_x(\sigma, \sigma') = \beta \left[\lambda U(x, \sigma) - \lambda U(x,\sigma') + W(\sigma, \sigma')\right]
\end{align}
where (for semi-reciprocity) we take the same potential $U(x, \sigma)$ corresponding to the force $F(x,\sigma)$, and with work $W(\sigma, \sigma') = - W(\sigma', \sigma)$ done by the chemical driving, allowing nonequilibrium features. 
We assume that for fixed $x$, the rates violate detailed balance, which, keeping equal temperature for the total system (probe and medium), is only possible for $n \geq 3$, \cite{Lucaactivesolid, entropyproductionsolvable}: 
 between at least three chemical states,
\begin{equation}
\frac{k_{x}(0,1)\,k_{x}(1,2)\,k_{x}(2,0)}
   {k_{x}(1,0) \, k_{x}(0,2) \,k_{x}(2,1)}
= e^{\beta \Delta\mu} \neq 1
\label{eq:cycle_affinity}
\end{equation}
 with $\Delta \mu$ the chemical affinity that drives each jumper out
of equilibrium. Hence, for at least three chemical states,
\begin{align}\label{dap}
  W(0, 1) + W(1,2) + W(2,0) = \Delta \mu 
\end{align}
In the simplest driven case, $n=3$, we take 
\begin{equation}\label{definition-w}
W(0,1) = W(1,2) = W(2,0) = w
\end{equation}
so that $\Delta \mu = 3 w$. This parameter $w$ plays an essential role in creating a nonequilibrium steady state for the bath and, as we will see, allows for rich behavior for the coupled probe. Note that we assume that the driving $W$ does not depend on $x$. Physically, that means that the (mechanical) probe does not directly interfere with the nonequilibrium driving on the (chemical) $\sigma$-medium. However, the probe does enter for determining the reactivities in \eqref{rate decomposition}; we assume the form 
\begin{align}\label{rela}
  \psi_x(\sigma, \sigma') = \psi_0 \left(1 + \lambda \,\Psi_x(\sigma, \sigma') \right) \geq 0
\end{align}
where $\Psi_x(\sigma, \sigma') = \Psi_x(\sigma', \sigma)$ will become essential. These reactivities can also depend on the driving $w$ or $\Delta\mu$, since a large driving field can cause trapping; see also \cite{Baerts_2013}. That will be denoted by $\psi_x(\sigma, \sigma'; w)$ when important, but otherwise is left out.\\

The stationary distribution $\rho_x(\sigma)$ of a single
jumper for the dynamics at fixed probe position $x(t) = x$ satisfies the stationary condition, \cite{kamp},
\begin{align}\label{bos}
  0 = \mathcal{L}_\sigma^\dagger \rho_x(\sigma) = \sum_{\sigma' \in \mathbb{Z}_n} \left[ k_x(\sigma, \sigma') \rho_x(\sigma) - k_x(\sigma', \sigma) \rho_x(\sigma') \right]
\end{align}
with $\mathcal{L}_\sigma^\dagger $ the forward generator of the Markov jump process for the jumpers at fixed $x$. This $\rho_x(\sigma)$ is the so-called Born-Oppenheimer or pinned distribution \cite{Born1927, szabo1996modern}. Under \eqref{eq:cycle_affinity}, it carries a nonzero (chemical) current $0\rightarrow 1\rightarrow 2$ if $\Delta\mu >0$, and we know that under strong enough driving $\Delta\mu$, the population distribution $\rho_x(\sigma)$ and the stationary current strongly depend on the behavior of $\Psi_x(\sigma, \sigma')$; see Section \ref{remark large driving} and \cite{heatb}.

\section{Reduced dynamics}\label{section reduced dynamics}
In biophysical systems, activity in general and oriented locomotion more specifically are not imposed directly but emerge from the underlying chemistry. To understand that scenario, we are interested in the reduced probe dynamics when integrating out the nonequilibrium jumper bath from the dynamics \eqref{eq:probe_dyn}. There exist standard techniques to do this, \textit{e.g.} projection operators \cite{Mori1, Mori2, zwanzig, Schilling,widder2026generalised, grabert1982projection}, nonequilibrium linear response theory and its path-space action \cite{maesresponse, frenesy, over, under, Tanogami2022} or singular perturbation theory \cite{singularperturbationtheory, singularescaperate, Tanogami2022, Tikhonov1952, Lomov1992}. These methods require a time-scale separation in the sense that both mechanical variables $v(t)$ and $x(t)$ are slow variables compared to the chemical processes. References where statistical mechanics meet mechanobiology include \cite{Bravi_2017,Polettini_2017,Bo_2017} for using projection methods to study effective thermodynamic evolution in systems with hidden degrees of freedom, while 
\cite{Seifert2011} gives a review that bridges mechanochemical coupling and stochastic dynamics. Following these techniques, 
in the Markovian limit, the final result is an effective or reduced diffusion process for the probe of the form\footnote{This equation is equivalent to the following Fokker-Planck equation for the reduced distribution $\rho_{\text{probe}}(x,v,t) = \sum_{\sigma \in \mathbb{Z}_n} \rho_{\text{tot}}(x,v,\sigma,t)$
\begin{align*}
 \frac{\partial \rho_{\text{probe}}}{\partial t}(x,v,t) &=-  \frac{\partial}{\partial x} \left[\rho_{\text{probe}}(x,v,t) \ v \right] - \frac{\partial}{\partial v} \left[ \rho_{\text{probe}}(x,v,t) \ \frac{1}{m}\left(- \frac{\partial V}{\partial x}(x) + \lambda \bar{F}(x)\right) \right]\\
  & \qquad - \frac{\partial}{\partial v} \left[\rho_{\text{probe}}(x,v,t) \left( -\frac{\nu(x)}{m} \ v \right) \right] +\frac{\partial^2}{\partial v^2} \left[ \rho_{\text{probe}}(x,v,t) \ \frac{B(x)}{m^2} \right] \nonumber
\end{align*}}, 
\begin{align}
 \dot{x}(t) &= v(t), \qquad m\dot{v}(t) = -\frac{\partial V}{\partial x}(x(t)) + \lambda \bar F(x(t))- \nu(x(t)) \, v(t) + \sqrt{2 B(x(t))} \,\xi(t) \label{reduced dynamics}
\end{align}
with mean force/streaming term $\bar{F}(x)$, friction coefficient $\nu(x)$ and standard white noise $\xi(t)$, multiplied in It\^o-sense with the noise amplitude $B$. Taking the standard projection operator formulas, see \cite{PeiMaes2025, pei2026transferActiveMotion, beyen2025couplingelasticstringactive}, these induced terms are given by
\begin{align}
  \bar{F}(x) & = N \left \langle F(x, \sigma) \right \rangle_x^{\text{BO}} \label{mfo} \\
  \nu(x) & = - \lambda N \int_0^{\infty} \id \tau \ \left \langle \frac{\partial U}{\partial x}(x, \sigma(\tau)) \ ; \  \frac{\partial \log \rho_x}{\partial x}(\sigma) \right \rangle_x^{\text{BO}} \label{reduced form nu s} \\
  B(x) & = \lambda^2 N \int_0^{\infty} \id \tau \ \left \langle \frac{\partial U}{\partial x}(x, \sigma(\tau)) \ ; \ \frac{\partial U}{\partial x}(x, \sigma) \right\rangle_x^{\text{BO}} \label{reduced form D 2}
\end{align}
in terms of the Born-Oppenheimer distribution, which is the stationary distribution $\rho_x(\sigma)$ over the chemical degrees of freedom $\sigma$ for the dynamics defined under \eqref{rate decomposition} at fixed $x$. For example,
\begin{align*}
 \langle f\rangle_{x}^{BO}&= \sum_{\sigma \in \mathbb{Z}_n} f(x,\sigma) \ \rho_x(\sigma) \\
 C_x(\tau) &=\left \langle \frac{\partial U}{\partial x}(x, \sigma(\tau)) \ ; \ \frac{\partial U}{\partial x}(x, \sigma) \right\rangle_x^{\text{BO}} = \left \langle \frac{\partial U}{\partial x}(x, \sigma(\tau)) \, \frac{\partial U}{\partial x}(x, \sigma) \right\rangle_x^{\text{BO}} - \left(\left\langle \frac{\partial U}{\partial x}(x, \sigma) \right\rangle_x^{\text{BO}}\right)^2
\end{align*}
yields the time-covariance $C_x(\tau)$. 
It follows from stationarity that the noise amplitude is always positive since
\begin{align*}
  B(x) = \lambda^2 N \int_0^{\infty} \id \tau \ C_x(\tau) = \frac{\lambda^2 N}{2} \int_{- \infty}^{\infty} \id \tau \ C_x(\tau) = \frac{\lambda^2 N}{2} S_x(0)
\end{align*}
where $S_x(\omega) =\int_{- \infty}^{\infty} \id \tau \ C_x(\tau) \ e^{i \omega \tau } $ is the spectral density and $S_x(\omega) \geq 0$ by the Wiener–Khinchin theorem, \cite{kamp}. Moreover, it follows from \eqref{reduced form nu s}, that, for the study of reduced dynamics, the backreaction is essential; $\nu$ vanishes when the Born-Oppenheimer distribution $\rho(\sigma)$ is independent of $x$. Hence, if the bath dynamics is prescribed independently of the probe position, as in some simplified switching-potential models, the resulting reduced dynamics can be qualitatively different. \\

In equilibrium where the work function vanishes $W(\sigma, \sigma') = 0$, the Born-Oppenheimer distribution has the Boltzmann form $\rho_x(\sigma) = e^{- \beta \lambda U(x, \sigma)}/Z_x$ such that the streaming term \eqref{mfo} is derived from the free energy $\lambda \bar{F}(x) = - \partial_x \cal F(x), \ \cal F(x) = -Nk_B T \log Z_x$  and \eqref{reduced form nu s}--\eqref{reduced form D 2} satisfy
\begin{align}
  \nu(x)
  & = \beta \lambda^2 N \int_0^{\infty} \id \tau \ \left \langle \frac{\partial U}{\partial x}(x, \sigma(\tau)) \ ; \   \frac{\partial U}{\partial x}(x, \sigma) \right \rangle_x^{\text{BO}} \nonumber \\
  & = \beta B(x) \label{fdr2}
\end{align}
That equality indeed confirms the second fluctuation--dissipation theorem FDRII for equilibrium baths, \cite{balakrishnan2020elements,maes2014second}. In particular, the friction is always a positive quantity since $B(x)$ is. The main question of the present paper is to understand when and how all that changes when the chemical bath is driven, $\Delta\mu\neq 0$ in \eqref{eq:cycle_affinity}.\\

 For simplicity and to get explicit formul\ae, from now on we mostly work with a three-level system, $\sigma\in \mathbb Z_3$, even though the setup and the arguments hold for any $n \geq 3$. Moreover, to restrict the number of parameters, we take in \eqref{driv}, \eqref{rela} that
\begin{align}\label{only state 1}
U(x,\sigma) &= \delta_{\sigma, 0} \tilde{U}(x), \qquad 
\Psi_x(0,1) = \tilde{\Psi}(x) , \quad \Psi_x(1,2) = 0, \quad \Psi_x(2,0) = 0  
\end{align}
for some functions $\tilde{U}(x), \tilde{\Psi}(x)$ periodic in $[0,L)$. We also restrict our analysis to the weak coupling regime, considering leading order in $\lambda\ll 1$.

\subsection{Mean force}\label{mefo}
As derived in Appendix \ref{appendix calculation}, for the case \eqref{only state 1}, the mean force $\bar{F}(x) $ in \eqref{mfo} becomes 
 \begin{align}
  \lambda \bar{F}(x) &= -\frac{\lambda N}{3}\tilde{U}'(x) 
  +\lambda^{2} N \frac{(\cosh (\beta w)+1)}{3(2 \cosh (\beta w)+1)}\beta \tilde{U}(x) \ \tilde{U}'(x)
  \label{bar F weak} \\
  & \qquad +\lambda^2 N \frac{(3 \sinh (\beta w)+\cosh (\beta w)-1)}{9(2 \cosh (\beta w)+1)} \tilde{U}'(x) \tilde \Psi (x) 
+O(\lambda^3) \nonumber
\end{align}
with $w$ the driving from \eqref{definition-w}.
Then, by integrating over $x\in [0,L)$, we have 
\begin{align}
 \lambda\oint \id x \ \bar{F}(x) &= \lambda^2 N\frac{(3 \sinh (\beta w)+\cosh (\beta w)-1)}{9(2 \cosh (\beta w)+1)} \oint \id x \ \tilde{U}'(x) \tilde{\Psi}(x) \label{integral mean force 1 state}
\end{align}
 Although the original force $F$ in \eqref{eq:probe_dyn} is the gradient of a potential, a rotational part may appear in the mean force when \eqref{integral mean force 1 state} does not vanish, \cite{MaesNetocny2019, frenetic_Steering}. Note that the rotational part \eqref{integral mean force 1 state} of the mean force has a crucial dependence on the reactivities, which is an example of how time-symmetric kinetics becomes important in nonequilibrium, \cite{frenesy}. We also observe that the temperature-dependent prefactor has the same sign as the driving $w$, is asymmetric in $w \leftrightarrow -w$, and has finite (but different) limits for $w \to \pm \infty$.\\
As expected, the rotational part \eqref{integral mean force 1 state} always vanishes for the equilibrium case where $w = 0$, but more is true. Importantly, the mean force is always rotationless when $\tilde{\Psi}(x)$ does not depend on $x$ (due to the periodicity of $\tilde U(x)$). That also happens when the reactivities depend on $x$ only via the potential $U$, \textit{i.e.}, $\tilde{\Psi}(x) = \phi(\tilde{U}(x))$, because then \begin{equation*}
    \oint \id x \ \tilde{U}'(x)\psi(x) =\oint \id x \ \frac{\id \phi}{\id x}(\tilde{U}(x)) =0
  \end{equation*}
In these cases, we have an effective potential,
  $- \frac{\partial V}{\partial x}(x) + \lambda \bar{F}(x) = - \frac{\partial V_{\text{eff}}}{\partial x}(x) $, even in nonequilibrium. Clearly, we need some spatial phase difference between energy and the reactivities to generate a rotational force.\\

\begin{figure}[H]
  \centering
  \includegraphics[width=0.55\linewidth]{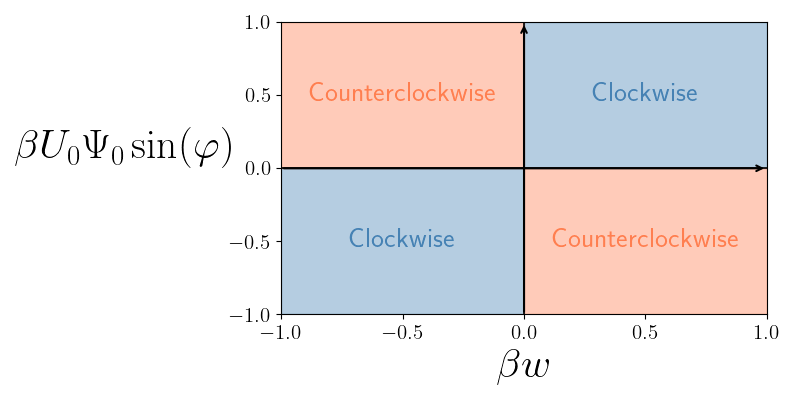}
  \caption{ Rotation direction of the mean force \eqref{integral mean force 1 state} in parameter space. We multiply by $\beta$ to obtain dimensionless quantities. } 
  \label{quadrantrotation}
\end{figure}

As a concrete yet illuminating example, consider the case where $\tilde{U}$ and $\tilde{\Psi}$ differ by a phase angle $\varphi$
  \begin{align}
  \tilde{U}(x) &= U_0 \cos\left(\frac{2 \pi }{L} x \right), \qquad 
  \tilde{\Psi}(x) = \Psi_0 \cos\left(\frac{2 \pi }{L} x + \varphi \right), \qquad \varphi \in (-\pi, \pi] \label{phase difference}
\end{align}
where $\Psi_0$ or $\varphi$ may depend on the driving $w$ and which delivers
\begin{align}\label{int sin phi}
  \oint \id x \ \tilde{U}'(x) \tilde{\Psi}(x) =U_0 \Psi_0 \,\pi \sin(\varphi) 
\end{align}
Hence, the force is rotational when there is a phase difference between the functions $\tilde{U}$ and $ \tilde{\Psi}$ and is maximal for $\varphi = \pi/2$, in which case $\tilde{U}'(x)$ and $\tilde{\Psi}(x)$ are in phase. 
That observation can be connected with the Rayleigh criterion (originally for thermo-acoustic instability), \cite{Rayleigh1878Explanation}: indeed, $\tilde{U}'$ plays the role of pressure, which must have the same phase as the reactivity $ \tilde{\Psi}$ to maximize the rotational push on the probe. \\
 Finally, whether the rotation is clockwise ($\lambda\oint \id x \ \bar{F}(x) > 0$) or counter-clockwise ($\lambda\oint \id x \ \bar{F}(x) < 0$) depends on the signs of $\beta w$ and $U_0 \Psi_0 \sin(\varphi)$, as depicted in Fig. \ref{quadrantrotation}. 

\subsection{Noise amplitude}
The noise amplitude in \eqref{reduced form D 2} is computed in Appendix \ref{appendix calculation}. For a 3-state bath with assumptions \eqref{only state 1}, we find 
\begin{align}
  B(x) & = \frac{2 \lambda^2 N \cosh \left(\frac{\beta w}{2}\right)}{9 \psi_0 (2 \cosh (\beta w)+1)} \tilde{U}'(x)^2 + O(\lambda^3) \label{B weak coupling result}
\end{align}
which is always positive and independent of the reactivity $\tilde{\Psi}(x)$. Moreover, it is symmetric in the driving $w$ and decreases with $|w|$, \textit{i.e.}, attains its maximum at equilibrium, $w = 0$.
Hence, the nonequilibrium corrections to $B(x)$ start from second order in $w$ only, in agreement with general results about the McLennan ensemble, \cite{McLennan, Khodabandehlou_2024}, while 
at large driving $w \to \pm\infty $, the noise vanishes, $  \lim_{w \to \pm \infty} B(x) = 0$; see also section \ref{remark large driving} and appendix \ref{appendix calculation}. That limit implies that the mechanical signal-to-noise ratio gets strongly enhanced by coupling with a large chemical driving, which living systems can use for enhanced stability and more reliable sensing, \cite{tuningtransduction, noneqsensing}.

\subsection{Induced linear drag}\label{subsect-friction-discussion}
As derived in Appendix \ref{appendix calculation} for the scenario \eqref{only state 1}, we obtain the friction coefficient \eqref{reduced form nu s} as,
\begin{align}
  \nu(x) & = \frac{\lambda^2 N}{9 \psi_0 (2 \cosh (\beta w)+1)^2} \Bigg[2 \beta \cosh \left(\frac{\beta w}{2}\right) (\cosh (\beta w)+2) \ \tilde{U}'(x)^2
  \label{frictionCoeff-WCoup} \,
  \\& \qquad \qquad \qquad \qquad\qquad\qquad \qquad  + e^{-\frac{\beta w}{2}} \left(e^{2\beta w}+ e^{ \beta w}-2\right) \ \tilde{U}'(x) \ \tilde{\Psi}'(x) \Bigg] + O(\lambda^3) \nonumber 
\end{align}
A natural decomposition follows 
\begin{equation}
  \nu(x) = \beta_{\text{eff}}\, B(x) + \gamma(x) + O(\lambda^3)
\end{equation}
with the effective inverse temperature $ \beta_{\text{eff}} := \beta\,\frac{2 + \cosh(\beta w) }{1 + 2 \cosh(\beta w)} \leq \beta $ defined from FDRII as the proportionality factor with $B(x)$ from \eqref{B weak coupling result}
and 
\begin{equation*}
  \gamma(x) = \lambda^2 N \frac{2 \sinh \left(\frac{\beta w}{2}\right) (e^{\beta w}+2) }{9 \psi_0 (2 \cosh (\beta w)+1)^2} \ \tilde{U}'(x) \ \tilde{\Psi}'(x) 
\end{equation*}
The term $\gamma(x)$ is the \emph{frenetic} contribution to the friction $\nu(x)$, reflecting its origin in the time-symmetric sector of dynamical fluctuations due to the presence of the reactivity $\tilde{\Psi}(x)$; see also \cite{PeiMaes2025,beyen2025couplingelasticstringactive}. Notably, and in stark contrast to the noise amplitude $B(x)$, this $\gamma(x)$ may take negative values. The competition between the entropic friction, $ \nu_{\text{ent}} = \beta_{\text{eff}} B(x)\geq 0$, and the frenetic term $\nu_{\text{fren}}(x) = \gamma(x)$ can give rise to regimes in which $\nu(x)$ becomes non-positive for some, or even all, $x \in [0, L)$. When the reactivities do not depend on $w$, we have that close to equilibrium $\gamma(x) = O(w)$, and hence, $\nu(x) >0$ for small enough $w$, indicating that the entropic part dominates around equilibrium. Similarly, when $\tilde{\Psi}$ does not depend on $w$, the friction vanishes as $w \to \pm \infty$, indicating an inhibition of the dissipation into the chemical bath at large driving, \cite{Lucaactivesolid}. Interestingly, when the reactivities do depend on $w$, as in \cite{Baerts_2013}, such that either of the limits
\begin{align*}
  \lim_{w \to \pm \infty } \frac{2 \sinh \left(\frac{\beta w}{2}\right) (e^{\beta w}+2) }{9 (2 \cosh (\beta w)+1)^2} \tilde{\Psi}'(x; w) \, = c(x) \neq 0
\end{align*}
do not vanish, we can have (positive or negative) friction without noise. In this case, that requires $\tilde{\Psi}$ to grow without bound as $w \to \pm \infty$ since the rates $k_x$ are not bounded in these limits. We say more about the large driving limit in Section \ref{remark large driving} and Appendix \ref{appendix calculation}. \\

In analogy with the analysis in Section \ref{mefo} for the mean force, $\gamma(x)$ vanishes when the reactivities do not depend on $x$, in which case we recover the FDRII (hence positive friction) at effective inverse temperature $\beta_{\text{eff}} \leq \beta $. That $x-$dependence in the reactivities was also crucial for generating a rotational component in the mean force. \\ Furthermore, it follows from \eqref{frictionCoeff-WCoup} that, generically, for every $x \in [0,L)$, the friction $\nu$ is non-monotonic as a function of $w$. From the natural decomposition $\nu = \nu_{\text{ent}} + \nu_{\text{fren}}$ with $ \nu_{\text{ent}}(x) = \beta_{\text{eff}} B(x)$ and, $ \nu_{\text{fren}}(x) = \gamma(x) $, it readily follows that the entropic part $\nu_{\text{ent}}$ is even in $w$ and decreases monotonically with $|w|$, such that the non-monotonicity of $\nu$ originates from its frenetic part, $\nu_{\text{fren}}$, another example of the role of frenesy in nonequilibrium systems. Interestingly, similar non-monotonic behaviour has been observed in active solids as well, \cite{Lucaactivesolid}. \\

\begin{figure}[H]
  \centering
  \includegraphics[width=0.6\linewidth]{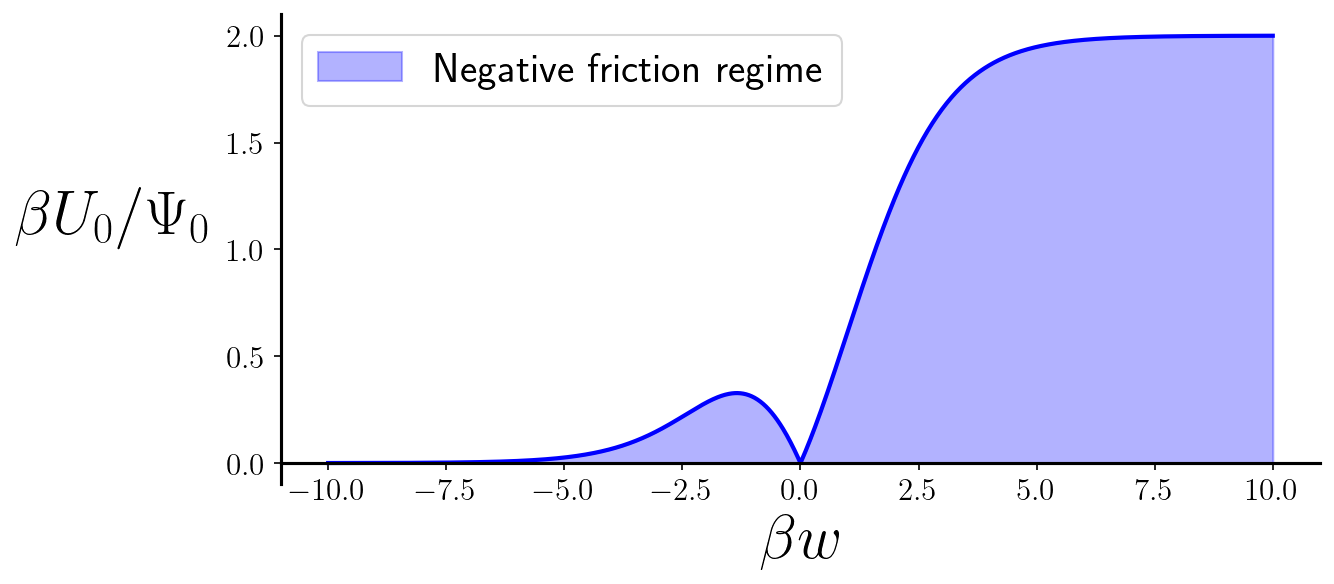}
  \caption{ Negative friction regime $\nu(x) \leq 0,$ for all $x \in [0,L)$ following \eqref{eq:negfric_negative_w}--\eqref{eq:negfric_positive_w} in parameter space. The blue curve corresponds to the equality and vanishes for $\beta w \to -\infty$, while it reaches $\beta U_0/\Psi_0 = 2$ as $\beta w \to \infty$. In the white region, $\nu(x) > 0$ for at least some $x \in [0,L)$. }
  \label{negativefrictiondiagram}
\end{figure}

\begin{figure}[H]
\centering
  \begin{subfigure}{0.50\linewidth}
    \centering
    \includegraphics[width=\linewidth]{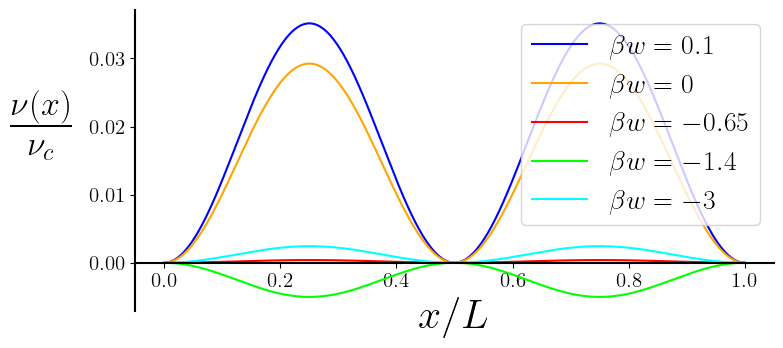}
    \caption{ Varying the driving $w$.}
    \label{nu for different w}
  \end{subfigure}
  \hfill 
  \begin{subfigure}{0.49\linewidth}
    \centering
    \includegraphics[width=\linewidth]{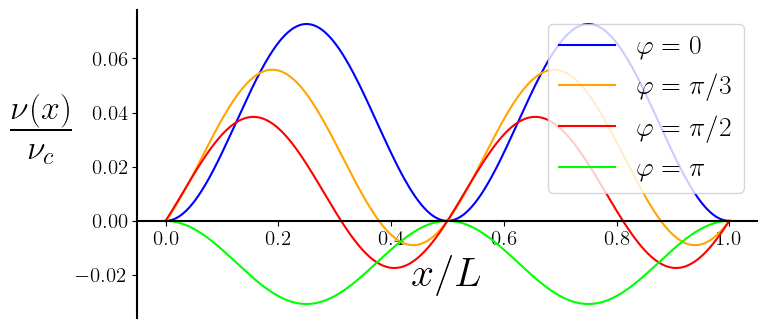}
    \caption{ Varying the phase angle $\varphi$.}
    \label{nu_negative_phase_angles}
  \end{subfigure}
  \caption{\raggedright Plot of the friction coefficient $\nu(x)/\nu_c$ with $\nu_c = 1/(\psi_0 \beta L^2)$ versus $x/L$. Parameter values: (a) $\Psi_0 = 4, \beta U_0 = 1, \varphi = 0, N = 100, \lambda = 0.1$ for several values of $\beta w$ and (b) $\Psi_0 = 4$, $\beta U_0 = 1, \beta w = 1, N = 100, \lambda = 0.1$ for different phase $\varphi$.}
  \label{combined_nu_plots}
\end{figure}

As a concrete example, to evaluate the sign of the linear friction coefficient $\nu$, we substitute functions \eqref{phase difference} in \eqref{frictionCoeff-WCoup}, yielding
\begin{align}
  \nu(x)=& \frac{4 \pi^2\lambda^2 N}{9(2 \cosh (\beta w)+1)^2} \frac{1}{\psi_0 \beta L^2} \Bigg[2 (\beta U_0)^2 \cosh \left(\frac{\beta w}{2}\right) (\cosh (\beta w)+2)\sin^{2}\left(\frac{2\pi}{L} x\right) \label{nu phase shift}
  \\& \qquad \qquad \qquad \qquad \qquad + e^{-\frac{\beta w}{2}} \left(e^{2\beta w}+ e^{ \beta w}-2\right) \beta U_0 \Psi_{0} \sin\left(\frac{2\pi}{L} x\right)\sin\left(\frac{2\pi}{L} x+\varphi\right)\Bigg] \nonumber 
\end{align}
Assuming (without loss of generality) that $U_{0},\Psi_{0}>0$, one obtains by a straightforward computation the following conditions on the sign of the friction, which are also depicted in Fig \eqref{negativefrictiondiagram}.
\begin{condenum}
\item $\nu(x)\leq 0$ for all $x \in [0,L) $, when $\varphi=0$ (in phase) and $w\in (-\infty,0)$, combined with 
    \begin{align}\label{eq:negfric_negative_w}
 \frac{\beta U_0}{\Psi_0} \leq \frac{\left(2 - e^{2\beta w} - e^{\beta w} \right)}{(1 + e^{\beta w})\!\left(\cosh(\beta w)+2\right) } 
    \end{align}
\item $\nu(x)\leq 0$ for all $x \in [0,L) $, when $\varphi=\pi$ (antiphase) and $w\in (0,\infty)$, combined with
    \begin{align}\label{eq:negfric_positive_w}
  \frac{\beta U_0}{\Psi_0} \leq \frac{\left( e^{2\beta w} + e^{\beta w}-2 \right)}{(1 + e^{\beta w})\!\left(\cosh(\beta w)+2\right) } 
    \end{align}
\end{condenum}
 The ratio $\beta U_0/\Psi_0$ serves as a measure for the strength of the entropic to frenetic part, such that the inequalities imply that this ratio cannot be too large for fully negative friction to occur, \textit{i.e.} the frenesy should be strong enough. 
When not (for $\varphi\in (-\pi,0)\cup (0,\pi)$), there is always (for all $w\in \mathbb{R}$) some $x\in [0,L)$ where the friction $\nu(x)>0$; it cannot be fully non-positive.
We refer to Appendix \ref{appendix calculation} for more details on the calculations; similar criteria can be derived for $\nu(x) \geq 0$.
We plot the function \eqref{nu phase shift} in Fig.~\ref{combined_nu_plots} and illustrate the conditions above for different parameter values. Fig. \ref{nu for different w} shows the influence of the dimensionless driving parameter $\beta w$ on the friction coefficient. Following \eqref{eq:negfric_negative_w}, the friction becomes fully negative when $-2.255 <\beta w < -0.666$. Indeed, slightly above the upper critical value, at $\beta w = -0.65$ (red curve), the friction almost vanishes $\nu(x) \approx 0$ for all $x/L$, while it becomes negative for smaller values (lime curve at $\beta w = -1.4)$ to become positive below the lower critical value (cyan curve at $\beta w = -3)$. That is a concrete example of the non-monotonic behavior of the friction in $w$, mentioned above. Fig.~\ref{nu_negative_phase_angles} shows the friction for different phase angles, giving examples where the friction is positive for some $x$ and negative for others. Following \eqref{eq:negfric_positive_w}, for $\varphi = \pi$ (lime curve), the friction becomes non-positive everywhere for $\beta w > 0.442$, and the choice $\beta w = 1$ indeed lies in that regime, while for $\varphi = 0$ the friction needs to be (partially) positive since $\beta w > 0$ and thus cannot satisfy \eqref{eq:negfric_negative_w}. 

\subsection{ From linear to saturation regime }

\begin{figure}[ht]
  \centering
  \begin{subfigure}[b]{0.49\textwidth}
    \centering
    \includegraphics[width=\textwidth]{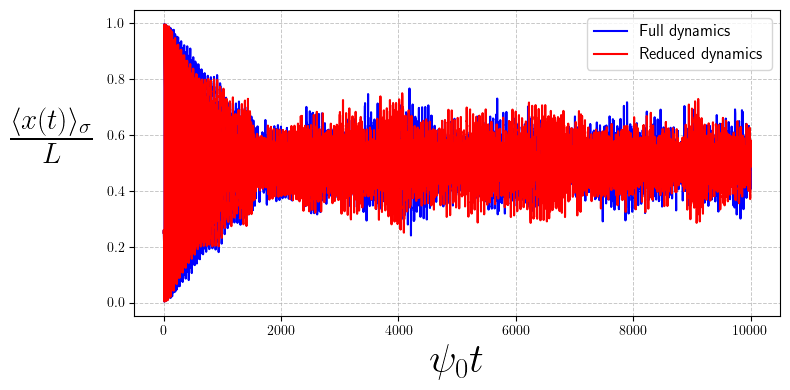}
    \caption{Average position }
    \label{reduced1}
  \end{subfigure}
  \begin{subfigure}[b]{0.49\textwidth}
    \centering
    \includegraphics[width=\textwidth]{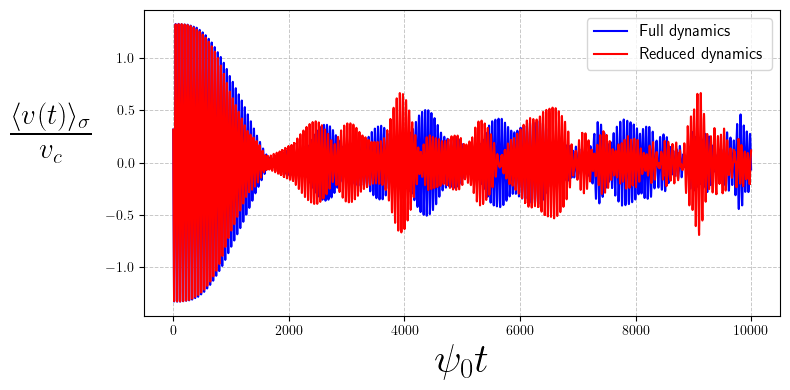}
    \caption{Average velocity }
    \label{reduced2}
  \end{subfigure}
  \vspace{1ex}
  \begin{subfigure}[b]{0.49\textwidth}
    \centering
    \includegraphics[width=\textwidth]{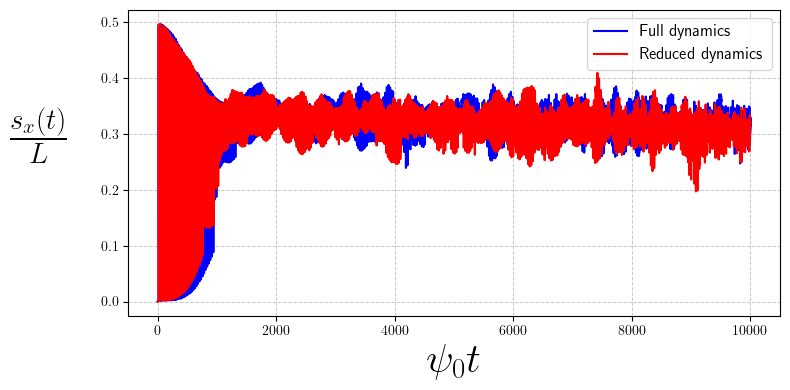}
    \caption{Position standard deviation }
    \label{reduced3}
  \end{subfigure}
  \begin{subfigure}[b]{0.49\textwidth}
    \centering
    \includegraphics[width=\textwidth]{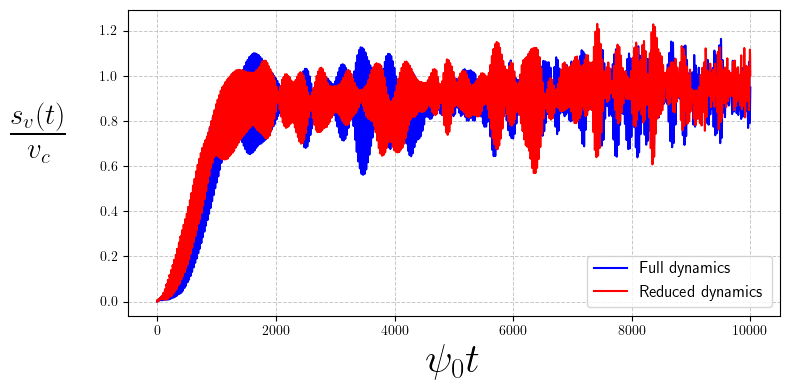}
    \caption{Velocity standard deviation }
    \label{reduced4}
  \end{subfigure}
  \caption{ \raggedright Position and velocity statistics in time for the full and reduced dynamics. Here $\langle \cdot \rangle_\sigma$ indicates an average over the $\sigma-$noise realizations, $s_x,s_v$ represent the standard deviations, while $v_c = \sqrt{|V_0|/m}$ and $L$ are characteristic scales. We used the parameters in Table \ref{table parameter values neg} for 50 trajectories with $x(0)/L = 0.25, \ \ v(0)/v_c = 0.316$. }
  \label{fig:hoofdfix2}
\end{figure}

As an illustration of the correctness of the reduced dynamics \eqref{reduced dynamics} in the time-scale separation regime, we compare it to the full setup \eqref{eq:probe_dyn} in Fig. \ref{reduced1}--\ref{reduced4}. They indicate that the reduced equation captures the essential features of the dynamics well, but it is not perfect; \textit{e.g.}, some signals get out of sync. Different parameter regimes also yield the same conclusion.

Upon elimination of the fast internal degrees of freedom, the probe may acquire rotational motion and/or 
a linear instability when, far enough from equilibrium, the frenetic component dominates the feedback and renders the linear friction negative (in the sense discussed in Section \ref{subsect-friction-discussion}). This negative friction reflects systematic energy injection from chemical cycles into the mechanical degree of freedom. This predicted behaviour also becomes clearly visible from the numerical simulations of the full dynamics in the next section. \\

At first sight, one might expect that the presence of negative friction is related to a rotational component of the mean force. However, in the setup where $\tilde{U}(x)$ and $\tilde{\Psi}(x)$ have a phase shift (cf.~\eqref{phase difference}) equal to $\varphi = 0,\pi$, the effective friction coefficient is non-positive for all $x$ when \eqref{eq:negfric_negative_w} resp. \eqref{eq:negfric_positive_w} are satisfied, while the mean force vanishes (see \eqref{int sin phi}).

In other words, negative friction is not strictly related to the creation of a net rotational current for the probe; it can only signify the emergence of a sustained activity with cyclic motion in $(x,v)$-space, similar to the limit cycle behavior of a Rayleigh oscillator
\begin{equation}\label{Rayleigh oscillator}
m \ddot x(t) + k x(t) =\gamma_1 \dot x(t) - \gamma_2 \dot x(t)^3 
\end{equation}
with $\gamma_1, \gamma_2 > 0$. That is the subject of the next section.\\ 

 Adding a damping term $- \nu_0 \dot x(t)$ with $\nu_0 > 0$ and corresponding thermal noise $\sqrt{2 \nu_0 k_B T} \eta(t)$ (following the fluctuation-dissipation relation) to \eqref{eq:probe_dyn}, results in the modified reduced dynamics
\begin{align*}
 m \ddot x(t) = m\dot{v}(t) &= -\frac{\partial V}{\partial x}(x(t)) + \lambda \bar F(x(t))- [\nu_0 + \nu(x(t))] \, v(t) + \sqrt{2 [\nu_0 k_B T + B(x(t))]} \,\xi(t) 
\end{align*}
with $\nu(x), B(x)$ from \eqref{reduced form nu s}--\eqref{reduced form D 2}. Then, the total friction coefficient becomes $\nu_{\text{tot}}(x) = \nu_0 + \nu(x)$ and the total noise amplitude is $B_{\text{tot}}(x) = \nu_0 k_B T + B(x)$. Since $\nu_0 = O(1)$ is generically larger than the induced term $\nu(x) = O(\lambda^2)$, the total friction $\nu_{\text{tot}}$ will be positive in general, and the Rayleigh instability is suppressed. However, the negativity of $\nu(x)$ does imply that the friction coefficient $\nu_{\text{tot}}(x)$ is reduced and becomes position dependent.
Nevertheless, the effective dynamics remains genuinely nonequilibrium because the fluctuation--dissipation relation remains violated, \textit{i.e.} $\nu_{\text{tot}} \neq k_B T B_{\text{tot}}$. Moreover, the mean force may still possess a rotational component such that the rotational regime survives. 

\section{Saturation regime}\label{section saturation regime}
The induced probe dynamics \eqref{reduced dynamics} has its limitations when the friction coefficient is negative. It implies an initial instability where the slow probe starts to accelerate, such that eventually the required time scale separation breaks down. That point is exactly the opportunity for getting an interesting (nonlinear) behavior or saturation regime. \\
We have simulated the coupled dynamics \eqref{eq:probe_dyn}--\eqref{rate decomposition} for the simple setup \eqref{only state 1}, \eqref{phase difference}, and compared these results with the most important consequences from the linear regime, like the presence of a rotational force and negative friction. We refer to Appendix \ref{appendix numerics} for the numerical implementation, while the code is available at \cite{numericalsimulation}. Table \ref{table different saturation regimes} gives an overview of the main possibilities and their most important characteristics. In this section, $\langle \cdot \rangle_\sigma$ indicates an average over the $\sigma-$noise realizations. 
\begin{table}[H]
\centering
\begin{tabular}{ |p{3cm}||p{3cm}|p{3cm}|p{5cm}|p{3cm}|}
 \hline
Regime& Average \newline $\left \langle v(t) \right \rangle_\sigma$ & Standard deviation $ s_v(t) $ & Stationary velocity distribution $\rho^{\text{stat}}(v)$ &Negative friction\\
 \hline
 Equilibrium  & $0$  & $\sqrt{k_BT/m}$ & Maxwellian & No\\
 Active  & $0$  & $\neq \sqrt{k_BT/m}$ & Bimodal distribution& Yes\\
 Rotation&  $\neq 0$ & $\neq \sqrt{k_BT/m}$  & Asymmetric Bimodal or \newline shifted Maxwellian& Partially\\
 \hline
\end{tabular}
\caption{ Saturation regimes}
\label{table different saturation regimes}
\end{table}

\subsection{Active regime}\label{section active regime}

\begin{figure}[H]
  \centering
  \begin{subfigure}{0.495\linewidth}
    \centering
    \includegraphics[width=\linewidth]{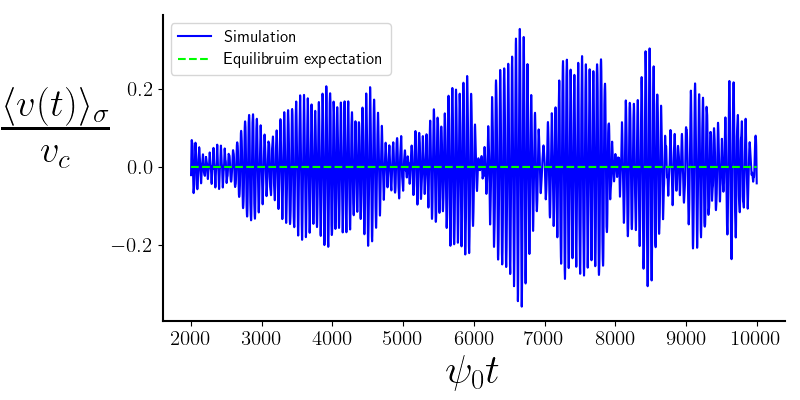}
    \caption{Average velocity}
    \label{avn}
  \end{subfigure}
  \hfill
  \begin{subfigure}{0.495\linewidth}
    \centering
    \includegraphics[width=\linewidth]{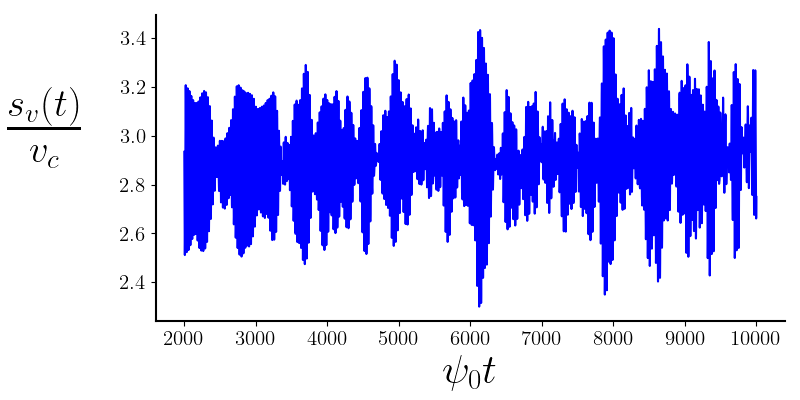}
    \caption{Velocity standard deviation}
     \label{stdvn}
  \end{subfigure}
  \caption{\raggedright Velocity statistics in time, (a) average velocity and (b) standard deviation, with characteristic scale $v_c = \sqrt{|V_0| /m}$ for 50 trajectories with $x(0)/L = 0.25, \ \ v(0)/v_c = 0.316$. We only plot the time interval $\psi_0 t \geq 2000$ for which the system has reached a steady state.}
  \label{combined_nesfric_plots}
\end{figure}

\begin{figure}[H]
  \centering
  \begin{subfigure}{0.495\linewidth}
    \centering
    \includegraphics[width=\linewidth]{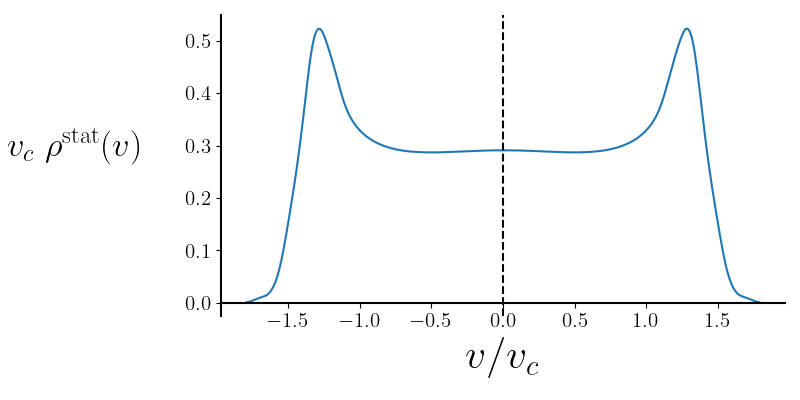}
    \caption{Velocity distribution}
    \label{rhovn1}
  \end{subfigure}
  \hfill
  \begin{subfigure}{0.495\linewidth}
    \centering
    \includegraphics[width=\linewidth]{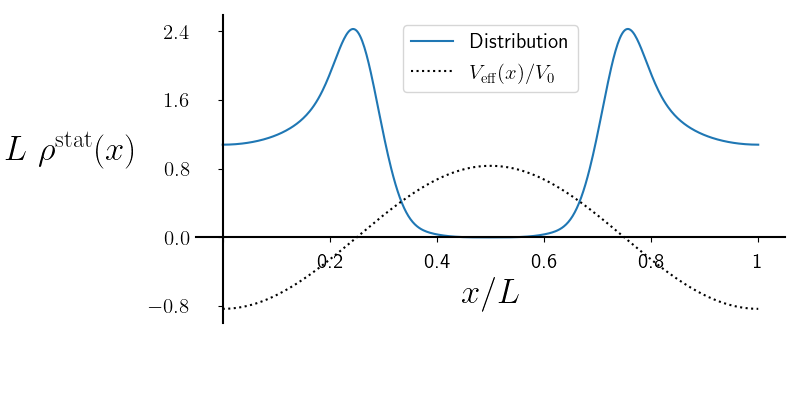}
    \caption{Position distribution}
    \label{rhoxn}
  \end{subfigure}
  \caption{\raggedright Steady-state distributions for the (a) velocity normalized with $v_c = \sqrt{|V_0| /m}$ and (b) position normalized with $L$. In (a), we added a dashed line around $v = 0$ for better visualization, while in (b), we added a dotted curve showing the effective potential $V_{\text{eff}}/V_0$ in which the probe moves. The distributions are obtained from the stationary data in Fig.~\ref{combined_nesfric_plots} using a Gaussian kernel density estimation.}
  \label{rhovn}
\end{figure}

The present paper addresses how to generate sustained mechanical activity (as exemplified in run-and-tumble processes modeling bacterial locomotion and other active particles \cite{activematterrev1, teVrugt2025, Gompper2020, activeparticle1}) from chemically driven stochastic dynamics, while maintaining a transparent structure of action, semi-reciprocal coupling, and local detailed balance.\\
We take the parameter values in Table \ref{table parameter values neg} to obtain negative friction with no rotational force (cf. \eqref{int sin phi}). That results in the stationary velocity statistics shown in Fig.~\ref{combined_nesfric_plots} and the stationary velocity distribution in Fig.~\ref{rhovn1}. In the steady regime, the mean velocity (in Fig.~\ref{avn}) fluctuates around zero, as in equilibrium. However, in contrast, there are frequent pulsations, an indication of the predicted negative linear friction when the probe is slow, after which the speed grows and saturates, to fall again to smaller values. The out-of-equilibrium feature can also be seen from the (normalized) velocity standard deviation in Fig.~\ref{stdvn}, which saturates around $\approx 2.9$, which is substantially different from the equipartition result, $ s_v^\text{eq} /v_c = 1/\sqrt{m k_B T v_c^2} = 0.316$. These features become especially apparent when looking at the stationary velocity distribution $\rho^{\text{stat}}(v)$ in Fig.~\ref{rhovn1}. Due to its (symmetrical) bimodal character, the mean value indeed vanishes, while the distance between the two peaks in velocity space roughly agrees with the stationary standard deviation. This bimodal structure is the clearest sign of the predicted negative linear friction since it indicates that the zero velocity state is a local minimum of the distribution, \textit{i.e.} it is unstable. Moreover, since the distribution is symmetric, there is no net current clockwise or counter-clockwise, indicating the lack of a rotational force, as expected. \\
We also plot the stationary position distribution $\rho^{\text{stat}}(x)$ in Fig.~\ref{rhoxn}, exhibiting a similar double-peaked symmetric structure. That can be understood as follows. Since there is no rotational part to the force, the probe moves under the influence of an effective potential $V_{\text{eff}}(x)$ derived from \eqref{bar F weak}. Keeping the periodic boundaries in mind, one observes that the distribution shows a $U-$shaped structure with peaks away from the minimum of $V_{\text{eff}}$, which is another characteristic property of activity, \cite{rtp_distribution}. We have thus generated chemo–mechanical activity in the (originally passive) Newtonian probe. From this perspective, it also explains why the distribution almost vanishes near $x/L = 1/2$, as the position distribution of active particles typically has a finite support. \\
Finally, the stationary probe distribution $\rho^{\text{stat}}(x,v)$ is given in phase space $(x,v)$ in Fig.~\ref{rho(x,v)}. Keeping the periodic boundaries into account, the distribution is strongly peaked around an elliptical region in this space, indicating that the probe is moving on a limit cycle akin to the Rayleigh oscillator \eqref{Rayleigh oscillator}. That behaviour is very different from the phase space of a pendulum moving under the influence of an induced positive friction due to the presence of the bath.
\begin{table}[H]
\centering
\caption{ Simulation parameters in active regime (arbitrary units).}
\begin{tabular}{lll}
\hline\hline
\textbf{Parameter} & \textbf{Symbol \ } & \textbf{Value} \\ 
\hline\hline
Ring length       & $L$       & 25 \\ 
Number of colloids    & $N$       & 100  \\ 
Coupling constant & $\lambda$ & $\frac{1}{\sqrt{N}} = 0.1$ \\ 
Mass     & $m$   & 1    \\ 
Inverse temperature     & $\beta$   & 1 \\
Amplitude $U$    & $U_0$       & 0.5   \\
Amplitude $V$    & $V_0$       & 10     \\
Amplitude $\psi$   & $\psi_0$       & 5    \\
Amplitude $\Psi$ & $\Psi_0$       & 5      \\
Driving    & $w$       & -1      \\
Phase angle    & $\varphi$       & 0    \\ 
\hline\hline
\label{table parameter values neg}
\end{tabular}
\end{table}

\begin{figure}[H]
  \centering
  \includegraphics[width=0.65\linewidth]{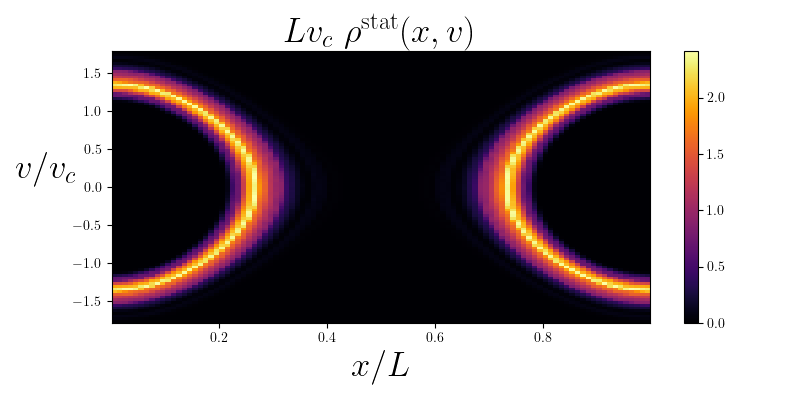}
  \caption{\raggedright Pseudocolor plot of the full steady state distribution $\rho^{\text{stat}}(x,v)$ normalized by the length $L$ and velocity $v_c$. It is peaked around an elliptical region, indicating limit cycle behaviour. The distribution is obtained from the histogram of the velocity data in Fig.~\ref{combined_nesfric_plots} as well as the position data (available from \cite{numericalsimulation}), subdivided into 100 bins.}
  \label{rho(x,v)}
\end{figure}

\subsection{Rotational regime}\label{section rotational regime}
\begin{figure}[H]
  \centering
  \begin{subfigure}{0.50\linewidth}
    \centering
    \includegraphics[width=\linewidth]{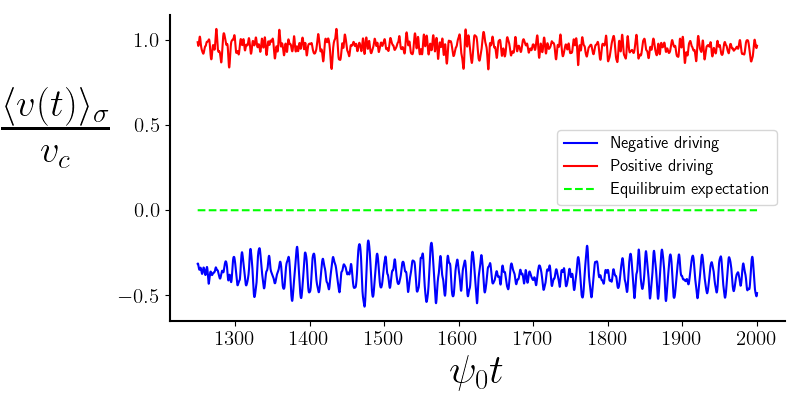}
    \caption{Average velocity}
    \label{avr}
  \end{subfigure}
  \hfill
  \begin{subfigure}{0.49\linewidth}
    \centering
    \includegraphics[width=\linewidth]{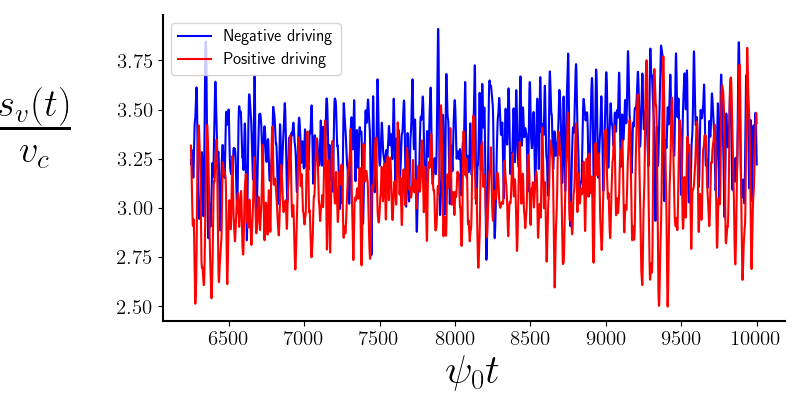}
    \caption{Velocity standard deviation}
     \label{stdvr}
  \end{subfigure}
  \caption{\raggedright Velocity statistics in time for positive and negative driving: (a) average velocity (b) standard deviation with characteristic scale $v_c = \sqrt{|V_0| /m}$ for 50 trajectories with $x(0)/L = 0.5, \quad v(0)/v_c = 0.316$. From $\psi_0 t \geq 1250$ onwards, the system has reached a steady state. The time axis has a different scale compared to Fig. \ref{combined_nesfric_plots} due to a different value of $\psi_0$.}
  \label{combined_rotfric_plots}
\end{figure}

\begin{figure}[H]
  \centering
  \includegraphics[width=0.65\linewidth]{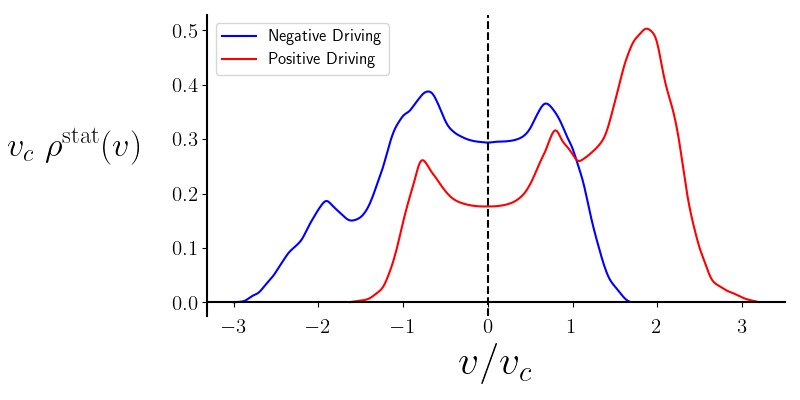}
  \caption{\raggedright Stationary velocity distributions for positive and negative driving with $v_c = \sqrt{|V_0| /m}$. As made explicit by the black dashed line around $v = 0$, there is a clear asymmetry between positive and negative velocities, indicating the presence of a rotational force. Reversing the sign of the driving also reverses the asymmetry. The distributions are obtained from the stationary velocity data in Fig. \ref{combined_rotfric_plots} using a Gaussian kernel density estimation.}
  \label{combined_disrotfric_plots}
\end{figure}

For the parameter values in Table \ref{table parameter values rot}, there is a rotational force according to \eqref{int sin phi}, while the friction takes on both positive and negative values as a function of $x \in [0, L)$. That results in the velocity statistics shown in Fig.~\ref{combined_rotfric_plots} and distributions in Fig.~\ref{combined_disrotfric_plots}. 
Contrary to the active regime above, the average velocity now saturates to a positive or negative value, indicating the presence of a current on the circle due to a rotational force. The sign of this average velocity agrees with \eqref{integral mean force 1 state}. For both positive and negative driving, the (normalised) standard deviation saturates around approximately the same value ($\approx 3$), ten times larger than the equilibrium value $ s_v^\text{eq} /v_c = 1/\sqrt{m k_B T v_c^2} = 0.316$. Moreover, the presence of this rotational force favors either positive or negative velocities and hence introduces an asymmetry in the bimodal velocity distribution as shown in Fig. \ref{combined_disrotfric_plots}. 
\begin{table}[H]
\centering
\begin{tabular}{lll}
\hline\hline
\textbf{Parameter} & \textbf{Symbol \ } & \textbf{Value} \\ 
\hline\hline
Ring length       & $L$       & 25     \\ 
Number of colloids    & $N$       & 100     \\ 
Coupling constant & $\lambda$ & $\frac{0.1}{\sqrt{N}} = 0.01$ \\ 
Mass     & $m$   & 1     \\ 
Inverse temperature     & $\beta$   & 1 \\
Amplitude $U$    & $U_0$       & 10    \\ 
Amplitude $V$    & $V_0$       & 10     \\ 
Amplitude $\psi$ & $\psi_0$       & 1      \\ 
Amplitude $\Psi$ & $\Psi_0$       & 5     \\ 
Driving    & $w$       & $\pm 2$     \\
Phase angle    & $\varphi$       & $ \frac{\pi}{4}$  \\
\hline\hline
\end{tabular}
\caption{Simulation parameters in the rotational regime (arbitrary units).}
\label{table parameter values rot}
\end{table}

\begin{figure}[H]
  \centering
  \includegraphics[width=0.65\linewidth]{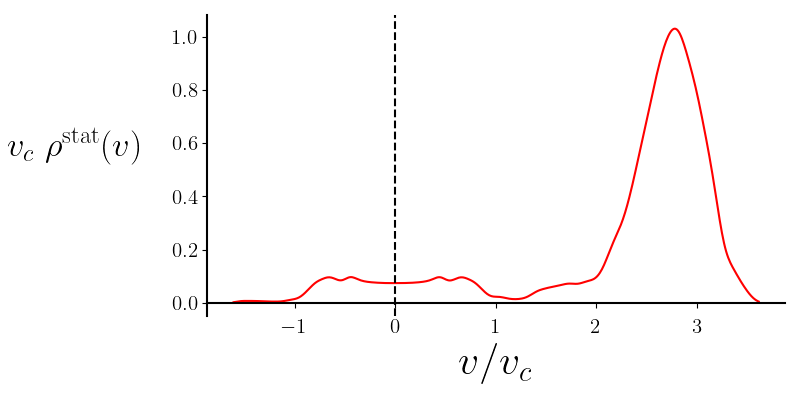}
  \caption{\raggedright The stationary velocity distribution at larger driving has a shifted Maxwellian form with $v_c = \sqrt{|V_0| /m}$ with a small peak around zero velocity. As made explicit by the black dashed line around $v = 0$, there is a strong asymmetry (due to the strong rotational mean force). The distribution is obtained from the stationary velocity data using a Gaussian kernel density estimation.}
  \label{shifted maxwellian}
\end{figure}

Upon increasing the driving $w$ to 2.5 (in arbitrary units) while keeping all other parameters unchanged, one expects the rotational force to increase (cf. \eqref{bar F weak}) and thus the asymmetry to become larger. That is indeed the case for Fig.~\ref{shifted maxwellian}, where one recognizes a shifted Maxwell distribution, with a small local maximum around zero velocity of small support.

\section{Additional remarks}\label{section remarks}
\subsection{Large driving limit}\label{remark large driving} 
\begin{figure}[H]
\centering
  \begin{subfigure}{0.50\linewidth}
    \centering
    \includegraphics[width=\linewidth]{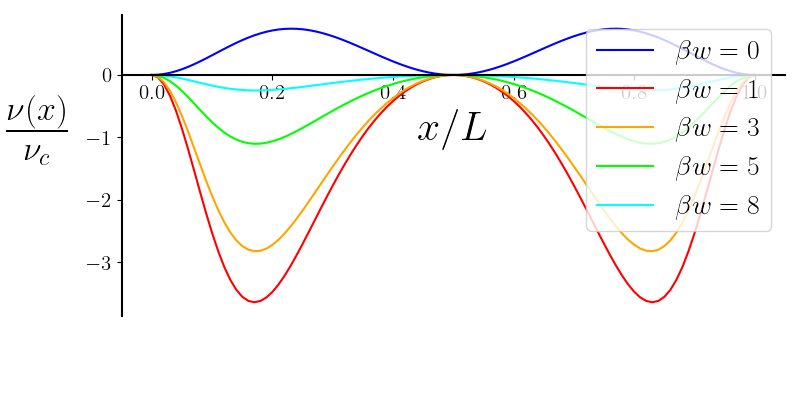}
    \caption{ Friction at increasing $w$. }
    \label{nu increasing w}
  \end{subfigure}
  \hfill 
  \begin{subfigure}{0.49\linewidth}
    \centering
    \includegraphics[width=\linewidth]{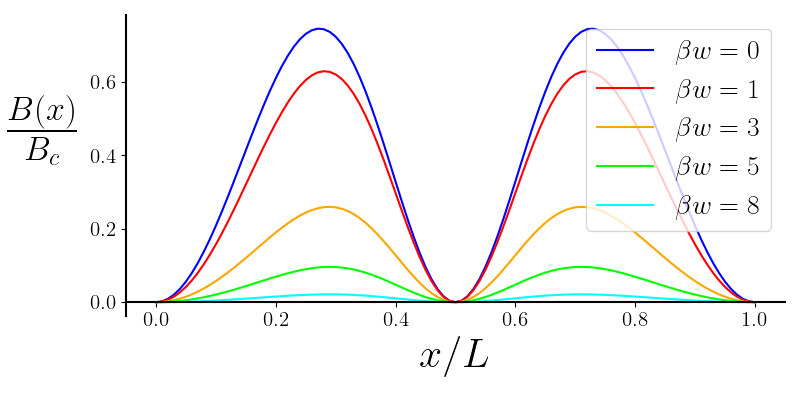}
    \caption{ Noise at increasing $w$. }
    \label{B increasing w}
  \end{subfigure}
  \caption{\raggedright Plot of (a) the friction coefficient $\nu(x)/\nu_c$ with $\nu_c = 1/(\psi_0 \beta L^2)$, and (b) the noise coefficient $B(x)/B_c$ with $B_c = 1/(\psi_0\beta^2 L^2)$, versus $x/L$ at increasing driving $\beta w$. We take the parameter values in \eqref{phase difference}, $\Psi_0 = 5, \beta U_0 = 0.5, \varphi = \pi$, and $N = 100, \lambda = 0.1$.}
  \label{figdifferentw}
\end{figure}

The influence of the chemical driving on the impulsive dynamics \eqref{eq:probe_dyn} obviously depends on the driving amplitude $w$ in \eqref{driv}--\eqref{definition-w}. Besides the equilibrium case $w = 0$, another limit of interest is large driving $w \to \pm \infty$. There, as derived in Appendix \ref{appendix calculation}, using the decomposition \eqref{rate decomposition}, the Born-Oppenheimer distribution \eqref{bos} satisfies
  \begin{align}\label{boslim}
\lim_{w \to \infty} \rho_x(\sigma) &\propto \, e^{\frac{\beta \lambda}{2} \left(U(x, \sigma + 1) - U(x, \sigma)\right)} \,\lim_{w \to \infty}\frac 1{\psi_x(\sigma,\sigma+1; w)} \\
\lim_{w \to - \infty} \rho_x(\sigma) &\propto \, e^{\frac{\beta \lambda}{2} \left(U(x, \sigma + 2) - U(x, \sigma)\right)} \,\lim_{w \to -\infty}\frac 1{\psi_x(\sigma+2,\sigma; w)} \nonumber 
\end{align}
where we allow for a possible $w-$dependence of the reactivities. These limits, keeping the directed energy differences and depending on the inverse reactivities, determine the Born-Oppenheimer distribution at large driving and essentially determine what chemical state $\sigma$ is dominating in the mean force $\bar{F}(x)$ \eqref{bar F weak} on the probe, \cite{heatb}. Given the coupling energy $U(x,\sigma)$, we can use the reactivities $\psi_x(\sigma,\sigma')$ to steer the probe motion; see also \cite{frenetic_Steering}. We refer to Appendix \ref{appendix calculation} for a calculation of the $w \to \pm \infty$ limit of the noise and friction. Assuming the limits $\lim_{w \to \pm\infty} \psi_x(\sigma, \sigma'; w)$ are finite and nonzero, it is shown there that the friction and noise vanish for $w \to \pm \infty$. That can also be deduced from plotting the formulas \eqref{reduced form nu s}--\eqref{reduced form D 2} for different $w$, yielding Fig. \ref{figdifferentw}. 
\subsection{Induced probe interaction}\label{induced probe interaction}
There is an increasing interest in nonreactive forces, especially in a biological context; see {\it e.g.}, \cite{metson2026emergentsinglespeciesnonreciprocitybistable}. We show how that emerges on the probe-level in the present context.\\
Suppose two independent probes $x_1, x_2$, each coupled to the chemical bath, and each subject to the same dynamics as in Section \ref{subsection model definition} with coupling potential $U(x_1, x_2, \sigma) = \delta_{\sigma,0} \left(\tilde{U}(x_1) + \tilde{U}(x_2) \right)$ so that there is no direct coupling between the probes $x_1, x_2$. 
Similar to \eqref{bar F weak}, we obtain the following induced mean forces on $x_1, x_2$, respectively,
\begin{align*}
  \lambda \bar{F}_1(x_1,x_2) & = - \lambda N \left \langle \frac{\partial U}{\partial x_1}(x_1, \sigma) \right \rangle_{x_1,x_2}^{\text{BO}} \\
  &= - \frac{\partial \mathcal{F}}{\partial x_1}(x_1,x_2) +\frac{(3 \sinh (\beta w)+\cosh (\beta w)-1)}{9(2 \cosh (\beta w)+1)} \frac{\partial \tilde{U}}{\partial x_1}(x_1) \tilde{\Psi}(x_1,x_2)\\
  \lambda \bar{F}_2(x_1,x_2) & =- \lambda N \left \langle \frac{\partial U}{\partial x_2}(x_2, \sigma) \right \rangle_{x_1,x_2}^{\text{BO}} \\
  &= - \frac{\partial \mathcal{F}}{\partial x_2}(x_1,x_2)+ \frac{(3 \sinh (\beta w)+\cosh (\beta w)-1)}{9(2 \cosh (\beta w)+1)} \frac{\partial \tilde{U}}{\partial x_2}(x_2) \tilde{\Psi}(x_1,x_2) \\
  \mathcal F(x_1,x_2) &= \frac{\lambda N}{3} \tilde{U}(x_1) - \frac{\lambda^2 N (e^{\beta w} + 1)}{12 (e^{2 \beta w} + e^{\beta w} + 1)} \beta\left(\tilde{U}(x_1) + \tilde{U}(x_2) \right)^2
\end{align*}
Consequently, the two probes become coupled through an effective potential $\mathcal F(x_1,x_2)$ (reciprocal interaction) as well as a non-potential (hence non-reciprocal) contribution proportional to the reactivities $\tilde{\Psi}$ since there are no $\partial_{x_1} \tilde{\Psi}, \partial_{x_2} \tilde{\Psi}$ terms. 
This is non-reciprocal in the sense that the Jacobian $J_{ij} = \frac{\partial \bar{F}_i}{\partial x_j}$ is not symmetric, \cite{nonreciprocalmanybodyphysics},
\begin{align*}
J_{12}-J_{21} & = \frac{\partial \bar{F}_1}{\partial x_2}(x_1,x_2) - \frac{\partial\bar{F}_2}{\partial x_1}(x_1,x_2) \\
&= \frac{(3 \sinh (\beta w)+\cosh (\beta w)-1)}{9(2 \cosh (\beta w)+1)} \left[ \frac{\partial \tilde{U}}{\partial x_1}(x_1) \frac{\partial \tilde{\Psi}}{\partial x_2}(x_1,x_2) - \frac{\partial \tilde{U}}{\partial x_2}(x_2) \frac{\partial \tilde{\Psi}}{\partial x_1}(x_1,x_2) \right]
\end{align*}
For instance, for the case \eqref{phase difference} with $\tilde{\Psi}(x_1,x_2) = \tilde{\Psi}(x_1) + \tilde{\Psi}(x_2)$,
\begin{align*}
  J_{12}-J_{21} = \frac{4 \pi^2 U_0 \Psi_0(3 \sinh (\beta w)+\cosh (\beta w)-1)}{9 L^2(2 \cosh (\beta w)+1)} \sin\left(\frac{2 \pi}{L}(x_1-x_2) \right) \sin(\varphi)
\end{align*}
which only vanishes when $\varphi = \pi \ell, \ \ell \in \mathbb{Z}$. \\
 Due to this induced coupling, one also wonders about correlations between the probes, and the possible emergence of synchronization, \cite{Chatzittofi2025, Agudo_Canalejo_2021, Chatzittofi_2023,majumdar2026dynamicallyemergentcorrelations}. We keep that for future investigation.

\subsection{Real line}\label{section real line}
The derivation of the reduced dynamics remains unchanged when the probe moves on the real line $x \in \mathbb{R}$ instead of the circle. In particular, the linear instability due to negative friction remains unchanged, and a bimodal velocity distribution can occur. The only difference is that the force is always rotationless on the real line (derivable from a potential). \\

\begin{figure}[H]
  \centering
  \begin{subfigure}{0.495\linewidth}
    \centering
    \includegraphics[width=\linewidth]{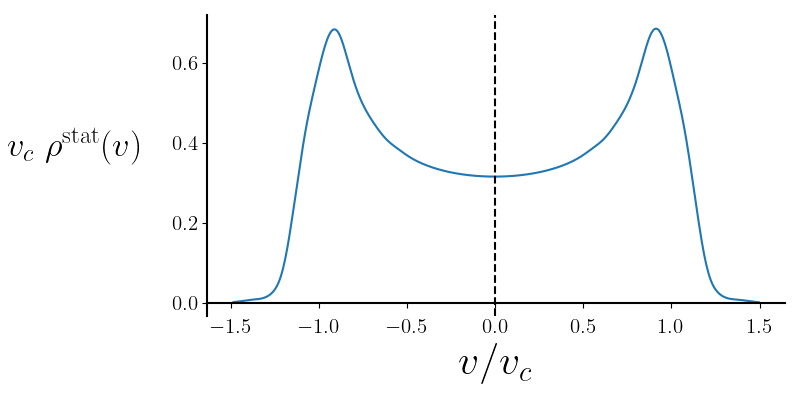}
    \caption{Velocity distribution}
    \label{linedvn}
  \end{subfigure}
  \hfill
  \begin{subfigure}{0.495\linewidth}
    \centering
    \includegraphics[width=\linewidth]{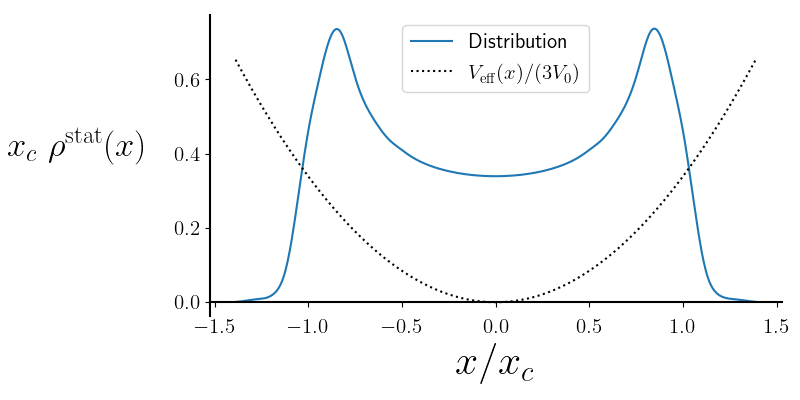}
    \caption{Position distribution}
     \label{linedxn}
  \end{subfigure}
  \caption{\raggedright Stationary (a) velocity and (b) position distributions for a probe on the line. In (a), we added a dashed line at $v = 0$ to highlight its bimodal character, while in (b) we also show the effective potential $V_{\text{eff}}(x)/(3V_0)$ with $V_0 = k_V x_c^2/2$, rescaled for visual comparison with the distribution.}
  \label{lined}
\end{figure}

As a proof of principle, consider the case
\begin{align}\label{potential real line}
  V(x) = \frac{k_V x^2}{2}, \qquad \tilde{U}(x) = \frac{k_U x^2}{2}, \qquad \tilde{\Psi}(x) = \Psi_0 e^{-(x/r)^2}
\end{align}
where the probe moves with an effective spring constant depending on the number $N_0(t)$ of jumpers in chemical state $0$. Under the free motion $\lambda = 0$, we recognize the typical scales $x_c = \sqrt{x_0^2 + \frac{m v_0^2}{k_V}}, \ v_c = \sqrt{v_0^2 + \frac{k_V x_0^2}{m}}$ with initial conditions $x_0,v_0$. \\
For the parameter values in Table \ref{table line}, one finds the distributions in Fig.~\ref{lined}. Similarly to the case on the ring, the velocity and position distributions have a bimodal form due to the presence of a negative linear friction. Moreover, since there is no rotational part to the force, the probe moves under the influence of an effective potential $V_{\text{eff}}(x)$ derived from \eqref{bar F weak}. As before, this $U$-shaped structure with peaks away from the minimum of $V_{\text{eff}}$ is a characteristic property of active particles moving in a potential, \cite{rtp_distribution}. 

\begin{table}[H]
\centering
\caption{Simulation parameters on the line (arbitrary units).}
\begin{tabular}{lll}
\hline\hline
\textbf{Parameter} & \textbf{Symbol \ } & \textbf{Value} \\ 
\hline\hline
Initial position       & $x_0$       & 2     \\
Initial velocity       & $v_0$       & 1     \\
Number of colloids    & $N$       & 100     \\ 
Coupling constant & $\lambda$ & $\frac{1}{\sqrt{N}} = 0.1$ \\
Mass     & $m$   & 1     \\ 
Inverse temperature     & $\beta$   & 1 \\ 
Spring constant $U$    & $k_U$       & 0.5    \\ 
Spring constant $V$    & $k_V$       & 10     \\
Amplitude $\psi$ & $\psi_0$       & 1      \\ 
Amplitude $\Psi$ & $\Psi_0$       & 5     \\
Driving    & $w$       & $0.5$     \\ 
Length scale $\Psi$    & $r$       & $ 1$  \\ 
\hline\hline
\label{table line}
\end{tabular}
\end{table}

\section{Conclusion}

The transfer of a low-entropy condition and associated nonequilibrium features from one set of degrees of freedom to another is a central question of nonequilibrium statistical mechanics with important relevance to biology. In the present paper, we have focused on theoretical modeling of chemo-mechanical transfer in a hybrid dynamical setup, and we found the emergence of activity and rotational motion. More specifically, we have shown that sustained mechanical activity can be generated for a Newtonian probe by coupling it to driven Markov jump processes,
while maintaining semi-reciprocity and local detailed balance with a single thermal bath. By activity for the probe on the ring, we mean here the occurrence of two sharp peaks in the otherwise symmetric velocity distribution.
The construction makes explicit that activity does not require
{\it ad hoc} nonconservative forces at the mechanical level.
Instead, it emerges from a thermodynamically consistent
chemo–mechanical coupling in which the internal jump dynamics
is driven by a nonzero chemical affinity.\\
By keeping the setup simple yet structurally faithful to local detailed balance, we sharpen the conceptual understanding of how nonequilibrium chemistry can generate mechanical motion, and under what precise conditions it does so.  More generally, that is an instance of the importance of the time-symmetric fluctuation sector, as summarized by the frenesy \cite{fren1,frenesy}, only possible thanks to the presence of net dissipation.
One finds that the friction has an entropic and frenetic component, with the latter enabling negative friction; see also \cite{PeiMaes2025,beyen2025couplingelasticstringactive}. The origin of activity is thus traced to the interplay between entropy production and dynamical activity (frenesy), rather than to explicit violation of mechanical laws. Interestingly, the criteria for the occurrence of rotational forces and for the emergence of negative linear friction are very similar to those derived by Rayleigh for the combustion instability. Also, the velocity and phase space distribution show the characteristics of a Rayleigh oscillator. \\

 There are many possible future perspectives of the current model. First, it is important to characterise the steady-state regime further, either numerically or analytically, and compare the results to the conjectured Rayleigh oscillator \eqref{Rayleigh oscillator}. Moreover, one wonders how robust the current setup is \textit{e.g.} when applied to higher dimensions or under the influence of additional friction and thermal fluctuations for the probe. Lastly, continuing the discussion in Section \eqref{induced probe interaction}, it is worth investigating whether strong correlations can occur between independent mechanical probes that interact with the same chemical environment, \cite{Chatzittofi2025, majumdar2026dynamicallyemergentcorrelations}. \\
Thinking about a chemo-mechanical coupling inside a washboard potential, similar to \cite{Chatzittofi2025, Agudo_Canalejo_2021, Chatzittofi_2023}, another interesting question is how to modify the Kramer escape rate for a probe coupled to a driven chemical bath.
 \\

{\bf Acknowledgments}\\
 We thank Marco Baesi for comments and useful suggestions. AB is supported by the Research Foundation-Flanders (FWO) doctoral fellowship 1152725N.

\bibliographystyle{unsrt} 
\bibliography{bib}

@article{maes2014second,
  author  = {C. Maes},
  title   = {{On the Second Fluctuation-Dissipation Theorem for Nonequilibrium Baths}},
  journal = {J. Stat. Phys.},
  volume  = {154},
  number  = {3},
  pages   = {705--722},
  year    = {2014},
  doi     = {10.1007/s10955-013-0907-5}
}

@article{MaesNetocny2019,
  author       = {Maes, C. and Netočný, K.},
  title        = {Nonequilibrium corrections to gradient flow},
  journal      = {Chaos},
  volume       = {29},
  number       = {7},
  pages        = {073109},
  year         = {2019},
  doi          = {10.1063/1.5098055}
}

@misc{nonreciprocalmanybodyphysics,
      title={Nonreciprocal many-body physics}, 
      author={Fruchart, M. and Vitelli, V.},
      year={2026},
      eprint={2602.11111},
      archivePrefix={arXiv},
      primaryClass={cond-mat.stat-mech},
      url={https://arxiv.org/abs/2602.11111}, 
      note = {arXiv:2602.11111 [cond-mat.stat-mech]}
}

@Article{ldb,
title={{Local detailed balance}},
author={C. Maes},
journal={SciPost Phys. Lect. Notes},
pages={32},
year={2021},
publisher={SciPost},
doi={10.21468/SciPostPhysLectNotes.32},
url={https://scipost.org/10.21468/SciPostPhysLectNotes.32},
}

@misc{vrugt2025exactlyactivematter,
      title={{What exactly is `active matter'?}}, 
      author={M. te Vrugt and B. Liebchen and M. E. Cates},
      year={2025},
      eprint={2507.21621},
      archivePrefix={arXiv},
      primaryClass={cond-mat.soft},
      url={https://arxiv.org/abs/2507.21621},
      note = {arXiv:2507.21621 [cond-mat.soft]}
}

@article{seifert1,
doi = {10.1088/1367-2630/14/10/103023},
url = {https://doi.org/10.1088/1367-2630/14/10/103023},
year = {2012},
publisher = {IOP Publishing},
volume = {14},
number = {10},
pages = {103023},
author = {Zimmermann, E. and Seifert, U.},
title = {{Efficiencies of a molecular motor: a generic hybrid model applied to the F1-ATPase}},
journal = {New J. Phys.}
}

@article{activematterrev1,
   author = "Ramaswamy, S.",
   title = "{The Mechanics and Statistics of Active Matter}", 
   journal= "Annu. Rev. Condens. Matter Phys.",
   year = "2010",
   volume = "1",
   number = "Volume 1, 2010",
   pages = "323-345",
   doi = "https://doi.org/10.1146/annurev-conmatphys-070909-104101",
   url = "https://www.annualreviews.org/content/journals/10.1146/annurev-conmatphys-070909-104101",
   publisher = "Annual Reviews",
   issn = "1947-5462",
   type = "Journal Article",
  }

@article{teVrugt2025,
  author    = {M. te Vrugt and R. Wittkowski},
  title     = {Metareview: a survey of active matter reviews},
  journal   = {Eur. Phys. J. E},
  year      = {2025},
  volume    = {48},
  number    = {2},
  pages     = {12},
  doi       = {10.1140/epje/s10189-024-00466-z},
  url       = {https://doi.org/10.1140/epje/s10189-024-00466-z},
  issn      = {1292-895X}
}

@article{activeparticle1,
title = {{The statistical physics of active matter: From self-catalytic colloids to living cells}},
journal = {Phys. A: Stat. Mech. Appl.},
volume = {504},
pages = {106-120},
year = {2018},
note = {Lecture Notes of the 14th International Summer School on Fundamental Problems in Statistical Physics},
issn = {0378-4371},
doi = {https://doi.org/10.1016/j.physa.2017.12.137},
url = {https://www.sciencedirect.com/science/article/pii/S0378437117313869},
author = {{\'E}. Fodor and M. C. Marchetti}
}

@article{activematter,
doi = {10.1088/1742-5468/aa6bc5},
url = {https://dx.doi.org/10.1088/1742-5468/aa6bc5},
year = {2017},
publisher = {IOP Publishing and SISSA},
volume = {2017},
number = {5},
pages = {054002},
author = {Ramaswamy, S.},
title = {Active matter},
journal = {J. Stat. Mech.: Theory Exp.}
}

@article{howfaractive,
   title={{How Far from Equilibrium Is Active Matter?}},
   volume={117},
   ISSN={1079-7114},
   url={http://dx.doi.org/10.1103/PhysRevLett.117.038103},
   DOI={10.1103/physrevlett.117.038103},
   number={3},
   journal={Phys. Rev. Lett.},
   publisher={American Physical Society (APS)},
   author={Fodor, É. and Nardini, C. and Cates, M. E. and Tailleur, J. and Visco, P. and van Wijland, F.},
   year={2016}}

@article{mmotors,
author = {Jülicher, F. and Ajdari, A. and Prost, J.},
year = {1997},
month = {10},
pages = {1269},
title = {Modeling molecular motors},
volume = {69},
journal = {Rev. Mod. Phys.},
doi = {10.1103/RevModPhys.69.1269}
}

@article{seifert2,
   title={Universal bound on the efficiency of molecular motors},
   volume={2016},
   ISSN={1742-5468},
   url={http://dx.doi.org/10.1088/1742-5468/2016/12/124004},
   DOI={10.1088/1742-5468/2016/12/124004},
   number={12},
   journal={J. Stat. Mech.: Theory Exp.},
   publisher={IOP Publishing},
   author={Pietzonka, P. and Barato, A. C. and Seifert, U.},
   year={2016}, pages={124004} }

@article{ParrondoDinis2007RatchetsParadoxicalGames,
  author    = {J. M. R. Parrondo and L. Din\'{i}s},
  title     = {Brownian motion and gambling: From ratchets to paradoxical games},
  journal   = {Contemporary Physics},
  volume    = {45},
  number    = {2},
  pages     = {147--157},
  year      = {2004},
  doi       = {10.1080/00107510310001644836}
}

@article{Parrondo1998ReversibleRatchets,
  author    = {J. M. R. Parrondo},
  title     = {Reversible ratchets as {Brownian} particles in an adiabatically changing periodic potential},
  journal   = {Phys. Rev. E},
  volume    = {57},
  number    = {6},
  pages     = {7297--7300},
  year      = {1998},
  doi       = {10.1103/PhysRevE.57.7297}
}

@article{heatb,
author = {C. Maes and K. Neto\v{c}n\'{y}},
title = {{Heat Bounds and the Blowtorch Theorem}},
journal = {Ann. H. Poincaré},
volume = {14},
number = {5},
pages = {1193--1202},
year = {2013},
}

@book{kamp,
author = {N. G. Van Kampen},
title = {Stochastic Processes in Physics and Chemistry},
publisher= { North Holland },
edition = {3rd},
year = {2007},
}

@article{Khodabandehlou_2024,
   title={Close-to-equilibrium heat capacity},
   volume={57},
   ISSN={1751-8121},
   url={http://dx.doi.org/10.1088/1751-8121/ad3ef2},
   DOI={10.1088/1751-8121/ad3ef2},
   number={20},
   journal={J. Phys. A: Math. Theor.},
   publisher={IOP Publishing},
   author={Khodabandehlou, F. and Maes, C.},
   year={2024}, pages={205001} }

@article{McLennan,
  title = {{Statistical Mechanics of the Steady State}},
  author = {McLennan, J. A.},
  journal = {Phys. Rev.},
  volume = {115},
  issue = {6},
  pages = {1405--1409},
  numpages = {0},
  year = {1959},
  publisher = {American Physical Society},
  doi = {10.1103/PhysRev.115.1405},
  url = {https://link.aps.org/doi/10.1103/PhysRev.115.1405}
}

@book{nondiss,
author = {C. Maes},
title = {Non-Dissipative Effects in Nonequilibrium Systems},
publisher = {Springer International Publishing},
year = {2018},
}

@article{ldb3,
author = {C. Maes and K. Netočný},
title = {Time-reversal and Entropy},
journal = {J. Stat. Phys.},
volume = {110},
pages = {269--310},
year = {2003}
}

@book{seif,
author    = {U. Seifert},
title     = {Stochastic Thermodynamics},
publisher = {Cambridge University Press},
year      = {2025}
}

@article{KatzLebowitzSpohn1984,
author  = {S. Katz and J. L. Lebowitz and H. Spohn},
title   = {Nonequilibrium steady states of stochastic lattice gas models of fast ionic conductors},
journal = {J. Stat. Phys.},
volume  = {34},
pages   = {497--537},
year    = {1984}
}

@article{quantumrayleigh,
  title = {{Quantum limit cycles and the Rayleigh and van der Pol oscillators}},
  author = {Arosh, L. B. and Cross, M. C. and Lifshitz, R.},
  journal = {Phys. Rev. Res.},
  volume = {3},
  issue = {1},
  pages = {013130},
  numpages = {18},
  year = {2021},
  publisher = {American Physical Society},
  doi = {10.1103/PhysRevResearch.3.013130},
  url = {https://link.aps.org/doi/10.1103/PhysRevResearch.3.013130}
}

@article{rtp_distribution,
  title = {{Run-and-tumble particle in one-dimensional confining potentials: Steady-state, relaxation, and first-passage properties}},
  author = {Dhar, A. and Kundu, A. and Majumdar, S. N. and Sabhapandit, S. and Schehr, G.},
  journal = {Phys. Rev. E},
  volume = {99},
  issue = {3},
  pages = {032132},
  numpages = {14},
  year = {2019},
  publisher = {American Physical Society},
  doi = {10.1103/PhysRevE.99.032132},
  url = {https://link.aps.org/doi/10.1103/PhysRevE.99.032132}
}

@article{Rayleigh01041883,
author = {Strutt (3rd Baron Rayleigh), J. W.},
title = {{XXXIII. On maintained vibrations}},
journal = {Philos. Mag.},
volume = {15},
number = {94},
pages = {229--235},
year = {1883},
publisher = {Taylor \& Francis},
doi = {10.1080/14786448308627342},


URL = { 
    
        https://doi.org/10.1080/14786448308627342
    
    

},
eprint = { 
    
        https://doi.org/10.1080/14786448308627342
    
    

}

}

@article{Chen_1994,
   title={{Renormalization Group Theory for Global Asymptotic Analysis}},
   volume={73},
   ISSN={0031-9007},
   url={http://dx.doi.org/10.1103/PhysRevLett.73.1311},
   DOI={10.1103/physrevlett.73.1311},
   number={10},
   journal={Phys. Rev. Lett.},
   publisher={American Physical Society (APS)},
   author={Chen, L. Y. and Goldenfeld, N. and Oono, Y.},
   year={1994}, pages={1311–1315} }

@article{frenesy,
title={{Frenesy: Time-symmetric dynamical activity in nonequilibria}},
volume={850},
ISSN={0370-1573},
url={http://dx.doi.org/10.1016/j.physrep.2020.01.002},
DOI={10.1016/j.physrep.2020.01.002},
journal={Phys. Rep.},
publisher={Elsevier BV},
author={Maes, C.},
year={2020}, pages={1–33} }

@article{beyen2025couplingelasticstringactive,
  title={Coupling an elastic string to an active bath: The emergence of inverse damping},
  author={Beyen, A. and Maes, C. and Pei, J.-H.},
  journal={Phys. Rev. E},
  volume={112},
  pages={L042103},
  year={2025},
  doi={10.1103/PhysRevE.112.L042103}
}

@article{Mori1,
  title = {{Statistical-Mechanical Theory of Transport in Fluids}},
  author = {Mori, H.},
  journal = {Phys. Rev.},
  volume = {112},
  issue = {6},
  pages = {1829--1842},
  numpages = {0},
  year = {1958},
  publisher = {American Physical Society},
  doi = {10.1103/PhysRev.112.1829},
  url = {https://link.aps.org/doi/10.1103/PhysRev.112.1829}
}

@book{grabert1982projection,
  title={Projection Operator Techniques in Nonequilibrium Statistical Mechanics},
  author={Grabert, H.},
  isbn={9783540116356},
  lccn={82007332},
  series={Communications and Control Engineering},
  url={https://books.google.be/books?id=PD1EAQAAIAAJ},
  year={1982},
  publisher={Springer-Verlag}
}

@article{Mori2,
    author = {Mori, H.},
    title = {{Transport, Collective Motion, and Brownian Motion}},
    journal = {Prog. Theor. Phys.},
    volume = {33},
    number = {3},
    pages = {423-455},
    year = {1965},
    month = {03},
    issn = {0033-068X},
    doi = {10.1143/PTP.33.423},
    url = {https://doi.org/10.1143/PTP.33.423},
    eprint = {https://academic.oup.com/ptp/article-pdf/33/3/423/5428510/33-3-423.pdf},
}

@article{Schilling,
   title={{On the generalized Langevin equation and the Mori projection operator technique}},
   volume={58},
   ISSN={1751-8121},
   url={http://dx.doi.org/10.1088/1751-8121/ae02cc},
   DOI={10.1088/1751-8121/ae02cc},
   number={40},
   journal={J. Phys. A: Math. Theor.},
   publisher={IOP Publishing},
   author={Widder, C. and Zimmer, J. and Schilling, T.},
   year={2025}, pages={405001} }

@book{balakrishnan2020elements,
  title={{Elements of Nonequilibrium Statistical Mechanics}},
  author={Balakrishnan, V.},
  isbn={9783030622336},
  url={https://books.google.be/books?id=_v0MEAAAQBAJ},
  year={2020},
  publisher={Springer International Publishing}
}

@article{zwanzig,
author = {Zwanzig, R.},
title = {{Ensemble Method in the Theory of Irreversibility}},
journal = {J. Chem. Phys.},
volume = {33},
number = {5},
pages = {1338-1341},
year = {1960},
issn = {0021-9606},
doi = {10.1063/1.1731409},
url = {https://doi.org/10.1063/1.1731409},
eprint = {https://pubs.aip.org/aip/jcp/article-pdf/33/5/1338/18820045/1338\_1\_online.pdf},
}

@article{maesresponse,
title={{Response Theory: A Trajectory-Based Approach}},
volume={8},
ISSN={2296-424X},
url={http://dx.doi.org/10.3389/fphy.2020.00229},
DOI={10.3389/fphy.2020.00229},
journal={Front. Phys.},
publisher={Frontiers Media SA},
author={Maes, C.},
year={2020}}

@article{M,
title={Macroscopic fluctuation theory},
volume={87},
ISSN={1539-0756},
url={http://dx.doi.org/10.1103/RevModPhys.87.593},
DOI={10.1103/revmodphys.87.593},
number={2},
journal={Rev. Mod. Phys.},
publisher={American Physical Society (APS)},
author={Bertini, L. and De Sole, A. and Gabrielli, D. and Jona-Lasinio, G. and Landim, C.},
year={2015}, pages={593–636} }

@article{F,
title = {{Fluctuations in Stationary Nonequilibrium States of Irreversible Processes}},
author = {Bertini, L. and De Sole, A. and Gabrielli, D. and Jona-Lasinio, G. and Landim, C.},
journal = {Phys. Rev. Lett.},
volume = {87},
issue = {4},
pages = {040601},
numpages = {4},
year = {2001},
publisher = {American Physical Society},
doi = {10.1103/PhysRevLett.87.040601},
url = {https://link.aps.org/doi/10.1103/PhysRevLett.87.040601}
}

@misc{hal,
author    = {H. Tasaki},
title     = {Two theorems that relate discrete stochastic processes to microscopic mechanics},
eprint    = {cond-mat/0706.1032v1},
archivePrefix = {arXiv},
year      = {2007},
note = {arXiv:0706.1032 [cond-mat.stat-mech]}
}

@book{Gaspard_2022, place={Cambridge}, title={{The Statistical Mechanics of Irreversible Phenomena}}, publisher={Cambridge University Press}, author={Gaspard, P.}, year={2022}}

@Article{fren1,
author    = {Basu, U. and Maes, C.},
journal   = {J. Phys. Conf. Ser.},
title     = {{Nonequilibrium Response and Frenesy}},
year      = {2015},
issn      = {1742-6596},
pages     = {012001},
volume    = {638},
doi       = {10.1088/1742-6596/638/1/012001},
publisher = {IOP Publishing},
}

@Article{under,
author    = {Baiesi, M. and Boksenbojm, E. and Maes, C. and Wynants, B.},
journal   = {J. Stat. Phys.},
title     = {{Nonequilibrium Linear Response for Markov Dynamics, II: Inertial Dynamics}},
year      = {2010},
issn      = {1572-9613},
number    = {3},
pages     = {492--505},
volume    = {139},
doi       = {10.1007/s10955-010-9951-6},
publisher = {Springer Science and Business Media LLC},
}

@article{singularescaperate,
    author = {Dygas, M. M. and Matkowsky, B. J. and Schuss, Z.},
    title = {{A singular perturbation approach to non‐Markovian escape rate problems with state dependent friction}},
    journal = {J. Chem. Phys.},
    volume = {84},
    number = {7},
    pages = {3731-3738},
    year = {1986},
    issn = {0021-9606},
    doi = {10.1063/1.450213},
    url = {https://doi.org/10.1063/1.450213},
    eprint = {https://pubs.aip.org/aip/jcp/article-pdf/84/7/3731/18958347/3731_1_online.pdf},
}

@article{Tikhonov1952,
  author    = {A. N. Tikhonov},
  title     = {Systems of differential equations containing small parameters in the derivatives},
  journal   = {Mat. Sb.},
  volume    = {31 (73)},
  number    = {3},
  year      = {1952},
  pages     = {575--586},
  note      = {in Russian},
  url       = {https://mi.mathnet.ru/eng/sm5548},
}

@book{Lomov1992,
  author    = {S. A. Lomov},
  title     = {Introduction to the General Theory of Singular Perturbations},
  series    = {Translations of Mathematical Monographs},
  volume    = {112},
  publisher = {American Mathematical Society},
  address   = {Providence, RI},
  year      = {1992},
  isbn      = {978-0-8218-4569-5}
}

@misc{numericalsimulation,
  author  = {A. Beyen},
  title   = {Inducing Activity by Chemo–Mechanical Coupling},
  url     = {https://github.com/AaronBeyen/Inducing-Activity-by-Chemo-Mechanical-Coupling},
  date    = {2026-03-08},
note = {\url{https://github.com/AaronBeyen/Inducing-Activity-by-Chemo-Mechanical-Coupling}}
}

@article{singularperturbationtheory,
   title={{Unattainability of Carnot efficiency in thermal motors: Coarse graining and entropy production of Feynman-Smoluchowski ratchets}},
   volume={98},
   ISSN={2470-0053},
   url={http://dx.doi.org/10.1103/PhysRevE.98.022102},
   DOI={10.1103/physreve.98.022102},
   number={2},
   journal={Phys. Rev. E},
   publisher={American Physical Society (APS)},
   author={Nakayama, Y. and Kawaguchi, K. and Nakagawa, N.},
   year={2018} }

@Article{over,
author    = {Baiesi, M. and Maes, C. and Wynants, B.},
journal   = {J. Stat. Phys.},
title     = {{Nonequilibrium Linear Response for Markov Dynamics, I: Jump Processes and Overdamped Diffusions}},
year      = {2009},
issn      = {1572-9613},
number    = {5–6},
pages     = {1094--1116},
volume    = {137},
doi       = {10.1007/s10955-009-9852-8},
publisher = {Springer Science and Business Media LLC},
}

@article{PeiMaes2025,
title        = {Induced friction on a probe moving in a nonequilibrium medium},
author       = {Pei, J.‑H. and Maes, C.},
journal      = {Phys. Rev. E},
volume       = {111},
number       = {3},
pages        = {L032101},
year         = {2025},
doi          = {10.1103/PhysRevE.111.L032101},
}

@article{pei2026transferActiveMotion,
   title={{Transfer of Active Motion from Medium to Probe via the Induced Friction and Noise}},
   volume={136},
   ISSN={1079-7114},
   url={http://dx.doi.org/10.1103/bwk3-3nmn},
   DOI={10.1103/bwk3-3nmn},
   number={3},
   journal={Phys. Rev. Lett.},
   publisher={American Physical Society (APS)},
   author={Pei, J.-H. and Maes, C.},
   year={2026}}

@Article{Gompper2020,
author    = {Gompper, G. and Winkler, R. G. and Speck, T. and Solon, A. and Nardini, C. and Peruani, F. and L\"owen, H. and others},
journal   = {J. Phys. Condens. Matter},
title     = {The 2020 motile active matter roadmap},
year      = {2020},
issn      = {1361-648X},
number    = {19},
pages     = {193001},
volume    = {32},
doi       = {10.1088/1361-648x/ab6348},
publisher = {IOP Publishing},
}

@Article{Tanogami2022,
author    = {Tanogami, T.},
journal   = {J. Stat. Phys.},
title     = {{Violation of the Second Fluctuation-dissipation Relation and Entropy Production in Nonequilibrium Medium}},
year      = {2022},
issn      = {1572-9613},
number    = {3},
volume    = {187},
doi       = {10.1007/s10955-022-02921-7},
publisher = {Springer Science and Business Media LLC},
}

@article{Wurthner2022BridgingScales,
   title={Bridging scales in a multiscale pattern-forming system},
   volume={119},
   ISSN={1091-6490},
   url={http://dx.doi.org/10.1073/pnas.2206888119},
   DOI={10.1073/pnas.2206888119},
   number={33},
  journal = {Proc. Natl. Acad. Sci. U.S.A.},
   publisher={Proceedings of the National Academy of Sciences},
   author  = {W{\"u}rthner, L. and Brauns, F. and Pawlik, G. 
             and Halatek, J. and Kerssemakers, J. and Dekker, C. 
             and Frey, E.},
   year={2022}}

@book{Sujith2021Thermoacoustic,
  author    = {Sujith, R. I. and Pawar, S. A.},
  title     = {Thermoacoustic Instability: A Complex Systems Perspective},
  publisher = {Springer},
  series    = {Springer Series in Synergetics},
  year      = {2021},
  doi       = {10.1007/978-3-030-81135-8}
}

@misc{widder2026generalised,
      title={{Generalised Langevin Dynamics: Significance and Limitations of the Projection Operator Formalism}}, 
      author={C. Widder and T. Schilling},
      year={2026},
      eprint={2604.20453},
      archivePrefix={arXiv},
      primaryClass={math-ph},
      url={https://arxiv.org/abs/2604.20453}, 
      note = {arXiv:2604.20453 [math-ph]}
}

@article{Rayleigh1878Explanation,
  Author = {Strutt (3rd Baron Rayleigh), J. W.},
  title   = {{The Explanation of Certain Acoustical Phenomena}},
  journal = {Proc. R. Inst. GB.},
  volume  = {8},
  pages   = {536--542},
  year    = {1878}
}

@article{Burkart2022ControlProteinPatterns,
  title   = {Control of protein-based pattern formation via guiding cues},
  author  = {Burkart, T. and Wigbers, M.C. and W{\"u}rthner, L. and Frey, E.},
  journal = {Nat. Rev. Phys.},
  volume  = {4},
  pages   = {511--527},
  year    = {2022},
  doi     = {10.1038/s42254-022-00461-3},
  publisher = {Nature Publishing Group}
}

@article{julicher1997modeling,
  author = {J\"ulicher, F. and Ajdari, A. and Prost, J.},
  title = {Modeling molecular motors},
  journal = {Rev. Mod. Phys.},
  volume = {69},
  number = {4},
  pages = {1269--1281},
  year = {1997},
  doi = {10.1103/RevModPhys.69.1269}
}

@article{Bravi_2017,
doi = {10.1088/1478-3975/aa7363},
url = {https://doi.org/10.1088/1478-3975/aa7363},
year = {2017},
publisher = {IOP Publishing},
volume = {14},
number = {4},
pages = {045010},
author = {Bravi, B. and Sollich, P.},
title = {Statistical physics approaches to subnetwork dynamics in biochemical systems},
journal = {Phys. Biol.}
}

@incollection{Sun2022BiochemoMech,
  author = {Sun, S.-Y. and Zhang, H.-X. and Fang, W. and Chen, X.-D. and Li, B. and Feng, X.-Q.},
  title = {Chapter Three - Bio-chemo-mechanical coupling models of soft biological materials: A review},
  editor = {Bordas, St{\'e}phane P.A.},
  booktitle = {Advances in Applied Mechanics},
  volume = {55},
  pages = {309--392},
  year = {2022},
  publisher = {Elsevier},
  issn = {0065-2156},
  doi = {10.1016/bs.aams.2022.05.004},
  url = {https://www.sciencedirect.com/science/article/pii/S0065215622000059}
}

@article{Feng2025MechanoChemoBio, title={Mechano-chemo-biological theory of cells and tissues: review and perspectives}, volume={41}, DOI={10.1007/s10409-025-25315-x}, number={7}, journal={Acta Mech. Sin.}, publisher={Springer Science and Business Media LLC}, author={Feng, X.-Q. and Li, B. and Lin, S.-Z. and Wang, M.-Y. and Chen, X.-D. and Zhang, H.-X. and Fang, W.}, year={2025}}

@article{frenetic_Steering,
   title={Frenetic steering: Nonequilibrium-enabled navigation},
   volume={34},
   ISSN={1089-7682},
   url={http://dx.doi.org/10.1063/5.0177223},
   DOI={10.1063/5.0177223},
   number={6},
   journal={Chaos},
   publisher={AIP Publishing},
   author={Lefebvre, B. and Maes, C.},
   year={2024}}

@article{Born1927,
  author    = {M. Born and R. Oppenheimer},
  title     = {{Zur Quantentheorie der Molekeln}},
  journal   = {Ann. Phys.},
  year      = {1927},
  volume    = {389},
  number    = {20},
  pages     = {457--484},
  doi       = {10.1002/andp.19273892002}
}

@book{szabo1996modern,
  title={{Modern Quantum Chemistry: Introduction to Advanced Electronic Structure Theory}},
  author={Szabo, A. and Ostlund, N.S.},
  isbn={9780486691862},
  lccn={lc96010775},
  series={Dover Books on Chemistry},
  url={https://books.google.be/books?id=k-DcCgAAQBAJ},
  year={1996},
  publisher={Dover Publications}
}

@misc{metson2026emergentsinglespeciesnonreciprocitybistable,
      title={Emergent single-species non-reciprocity from bistable chemical dynamics}, 
      author={J. Metson and R. Golestanian},
      year={2026},
      eprint={2603.21863},
      archivePrefix={arXiv},
      primaryClass={cond-mat.soft},
      url={https://arxiv.org/abs/2603.21863}, 
      note = {arXiv:2603.21863 [cond-mat.soft]}
}

@article{Bo_2017,
   title={{Multiple-scale stochastic processes: Decimation, averaging and beyond}},
   volume={670},
   ISSN={0370-1573},
   url={http://dx.doi.org/10.1016/j.physrep.2016.12.003},
   DOI={10.1016/j.physrep.2016.12.003},
   journal={Phys. Rep.},
   publisher={Elsevier BV},
   author={Bo, S. and Celani, A.},
   year={2017}, pages={1–59} }

@Article{Seifert2011,
author={Seifert, U.},
title={Stochastic thermodynamics of single enzymes and molecular motors},
journal={Eur. Phys. J. E},
year={2011},
day={15},
volume={34},
number={3},
pages={26},
issn={1292-895X},
doi={10.1140/epje/i2011-11026-7},
url={https://doi.org/10.1140/epje/i2011-11026-7}
}

@article{Polettini_2017,
   title={{Effective Thermodynamics for a Marginal Observer}},
   volume={119},
   ISSN={1079-7114},
   url={http://dx.doi.org/10.1103/PhysRevLett.119.240601},
   DOI={10.1103/physrevlett.119.240601},
   number={24},
   journal={Phys. Rev. Lett.},
   publisher={American Physical Society (APS)},
   author={Polettini, M. and Esposito, M.},
   year={2017}}

@article{Basu_2015,
   title={{How Statistical Forces Depend on the Thermodynamics and Kinetics of Driven Media}},
   volume={114},
   ISSN={1079-7114},
   url={http://dx.doi.org/10.1103/PhysRevLett.114.250601},
   DOI={10.1103/physrevlett.114.250601},
   number={25},
   journal={Phys. Rev. Lett.},
   publisher={American Physical Society (APS)},
   author={Basu, U. and Maes, C. and Netočný, K.},
   year={2015}}

@article{Baerts_2013,
   title={Frenetic origin of negative differential response},
   volume={88},
   ISSN={1550-2376},
   url={http://dx.doi.org/10.1103/PhysRevE.88.052109},
   DOI={10.1103/physreve.88.052109},
   number={5},
   journal={Phys. Rev. E},
   publisher={American Physical Society (APS)},
   author={Baerts, P. and Basu, U. and Maes, C. and Safaverdi, S.},
   year={2013}}

@article{tuningtransduction,
  title = {{Tuning Transduction from Hidden Observables to Optimize Information Harvesting}},
  author = {Nicoletti, G. and Busiello, D. M.},
  journal = {Phys. Rev. Lett.},
  volume = {133},
  issue = {15},
  pages = {158401},
  numpages = {6},
  year = {2024},
  publisher = {American Physical Society},
  doi = {10.1103/PhysRevLett.133.158401},
  url = {https://link.aps.org/doi/10.1103/PhysRevLett.133.158401}
}

@article{noneqsensing, title={Fundamental limits on nonequilibrium sensing}, volume={16}, DOI={10.1038/s41467-025-65058-7}, number={1}, journal={Nat. Commun.}, publisher={Springer Science and Business Media LLC}, author={Dechant, A. and Lutz, E.}, year={2025}}

@article{Fritz_2023,
   title={{Thermodynamically consistent model of an active Ornstein–Uhlenbeck particle}},
   volume={2023},
   ISSN={1742-5468},
   url={http://dx.doi.org/10.1088/1742-5468/acf70c},
   DOI={10.1088/1742-5468/acf70c},
   number={9},
   journal={J. Stat. Mech.: Theory Exp.},
   publisher={IOP Publishing},
   author={Fritz, J. H. and Seifert, U.},
   year={2023}, pages={093204} }

@article{Agranov_2024,
doi = {10.1088/1367-2630/ad4dd6},
url = {https://doi.org/10.1088/1367-2630/ad4dd6},
year = {2024},
publisher = {IOP Publishing},
volume = {26},
number = {6},
pages = {063006},
author = {Agranov, T. and Jack, R. L. and Cates, M. E. and Fodor, É.},
title = {Thermodynamically consistent flocking: from discontinuous to continuous transitions},
journal = {New J. Phys.}
}

@article{Agranov_2025,
doi = {10.1088/1367-2630/ae0c2e},
url = {https://doi.org/10.1088/1367-2630/ae0c2e},
year = {2025},
publisher = {IOP Publishing},
volume = {27},
number = {10},
pages = {104602},
author = {Agranov, T. and Jack, R. L. and Cates, M. E. and Fodor, É.},
title = {Entropy production rate in thermodynamically consistent flocks},
journal = {New J. Phys.}
}

@Article{Chatzittofi2025,
author={Chatzittofi, M. and Golestanian, R. and Agudo-Canalejo, J.},
title={Topological phase locking in stochastic oscillators},
journal={Nat. Commun.},
year={2025},
day={24},
volume={16},
number={1},
pages={4835},
issn={2041-1723},
doi={10.1038/s41467-025-60070-3},
url={https://doi.org/10.1038/s41467-025-60070-3}
}

@article{Lucaactivesolid,
  title = {Mechanical inhibition of dissipation in a thermodynamically consistent active solid},
  author = {Cocconi, L. and Chatzittofi, M. and Golestanian, R.},
  journal = {Phys. Rev. Res.},
  volume = {7},
  issue = {4},
  pages = {L042062},
  numpages = {8},
  year = {2025},
  publisher = {American Physical Society},
  doi = {10.1103/tgyv-5vqv},
  url = {https://link.aps.org/doi/10.1103/tgyv-5vqv}
}

@article{modellingmolecularmotors,
  title = {Modeling molecular motors},
  author = {J\"ulicher, F. and Ajdari, A. and Prost, J.},
  journal = {Rev. Mod. Phys.},
  volume = {69},
  issue = {4},
  pages = {1269--1282},
  numpages = {0},
  year = {1997},
  publisher = {American Physical Society},
  doi = {10.1103/RevModPhys.69.1269},
  url = {https://link.aps.org/doi/10.1103/RevModPhys.69.1269}
}

@article{FluxReversalinaSymmetricOpticalThermalRatchet,
  title = {{Observation of Flux Reversal in a Symmetric Optical Thermal Ratchet}},
  author = {Lee, S.-H. and Ladavac, K. and Polin, M. and Grier, D. G.},
  journal = {Phys. Rev. Lett.},
  volume = {94},
  issue = {11},
  pages = {110601},
  numpages = {4},
  year = {2005},
  publisher = {American Physical Society},
  doi = {10.1103/PhysRevLett.94.110601},
  url = {https://link.aps.org/doi/10.1103/PhysRevLett.94.110601}
}

@article{AsymmetricCycling,
  title = {{Asymmetric Cycling and Biased Movement of Brownian Particles in Fluctuating Symmetric Potentials}},
  author = {Chen, Y.-d.},
  journal = {Phys. Rev. Lett.},
  volume = {79},
  issue = {17},
  pages = {3117--3120},
  numpages = {0},
  year = {1997},
  publisher = {American Physical Society},
  doi = {10.1103/PhysRevLett.79.3117},
  url = {https://link.aps.org/doi/10.1103/PhysRevLett.79.3117}
}

@article{Molecularcombustionmotors,
  title = {Molecular combustion motors},
  author = {Magnasco, M. O.},
  journal = {Phys. Rev. Lett.},
  volume = {72},
  issue = {16},
  pages = {2656--2659},
  numpages = {0},
  year = {1994},
  publisher = {American Physical Society},
  doi = {10.1103/PhysRevLett.72.2656},
  url = {https://link.aps.org/doi/10.1103/PhysRevLett.72.2656}
}

@article{janusparticles,
    author = {Huang, M.-J. and Schofield, J. and Gaspard, P. and Kapral, R.},
    title = {{From single particle motion to collective dynamics in Janus motor systems}},
    journal = {J. Chem. Phys.},
    volume = {150},
    number = {12},
    pages = {124110},
    year = {2019},
    issn = {0021-9606},
    doi = {10.1063/1.5081820},
    url = {https://doi.org/10.1063/1.5081820},
    eprint = {https://pubs.aip.org/aip/jcp/article-pdf/doi/10.1063/1.5081820/15556675/124110_1_online.pdf},
}

@article{diffusionphoreticswimmer,
    author = {Sabass, B. and Seifert, U.},
    title = {Dynamics and efficiency of a self-propelled, diffusiophoretic swimmer},
    journal = {J. Chem. Phys.},
    volume = {136},
    number = {6},
    pages = {064508},
    year = {2012},
    issn = {0021-9606},
    doi = {10.1063/1.3681143},
    url = {https://doi.org/10.1063/1.3681143},
    eprint = {https://pubs.aip.org/aip/jcp/article-pdf/doi/10.1063/1.3681143/15448057/064508_1_online.pdf},
}

@article{stochasticmicroswimmer,
  title = {Entropy production and thermodynamic inference for stochastic microswimmers},
  author = {Chatzittofi, M. and Agudo-Canalejo, J. and Golestanian, R.},
  journal = {Phys. Rev. Res.},
  volume = {6},
  issue = {2},
  pages = {L022044},
  numpages = {7},
  year = {2024},
  publisher = {American Physical Society},
  doi = {10.1103/PhysRevResearch.6.L022044},
  url = {https://link.aps.org/doi/10.1103/PhysRevResearch.6.L022044}
}

@article{Cocconi_2025,
   title={Mechanical inhibition of dissipation in a thermodynamically consistent active solid},
   volume={7},
   ISSN={2643-1564},
   url={http://dx.doi.org/10.1103/tgyv-5vqv},
   DOI={10.1103/tgyv-5vqv},
   number={4},
   journal={Phys. Rev. Res.},
   publisher={American Physical Society (APS)},
   author={Cocconi, L. and Chatzittofi, M. and Golestanian, R.},
   year={2025}}

@misc{majumdar2026dynamicallyemergentcorrelations,
      title={{Dynamically Emergent Correlations}}, 
      author={S. N. Majumdar and G. Schehr},
      year={2026},
      eprint={2603.03162},
      archivePrefix={arXiv},
      primaryClass={cond-mat.stat-mech},
      url={https://arxiv.org/abs/2603.03162},
      note = {arXiv:2603.03162 [cond-mat.stat-mech]}
}

@article{Agudo_Canalejo_2021,
   title={{Synchronization and Enhanced Catalysis of Mechanically Coupled Enzymes}},
   volume={127},
   ISSN={1079-7114},
   url={http://dx.doi.org/10.1103/PhysRevLett.127.208103},
   DOI={10.1103/physrevlett.127.208103},
   number={20},
   journal={Phys. Rev. Lett.},
   publisher={American Physical Society (APS)},
   author={Agudo-Canalejo, J. and Adeleke-Larodo, T. and Illien, P. and Golestanian, R.},
   year={2021}}

@article{Chatzittofi_2023,
   title={Collective synchronization of dissipatively-coupled noise-activated processes},
   volume={25},
   ISSN={1367-2630},
   url={http://dx.doi.org/10.1088/1367-2630/acf2bc},
   DOI={10.1088/1367-2630/acf2bc},
   number={9},
   journal={New J. Phys.},
   publisher={IOP Publishing},
   author={Chatzittofi, M. and Golestanian, R. and Agudo-Canalejo, J.},
   year={2023}, pages={093014} }

@article{entropyproductionbrownianellipsoid,
  title = {{Entropy production of a Brownian ellipsoid in the overdamped limit}},
  author = {Marino, R. and Eichhorn, R. and Aurell, E.},
  journal = {Phys. Rev. E},
  volume = {93},
  issue = {1},
  pages = {012132},
  numpages = {15},
  year = {2016},
  publisher = {American Physical Society},
  doi = {10.1103/PhysRevE.93.012132},
  url = {https://link.aps.org/doi/10.1103/PhysRevE.93.012132}
}

@Article{Sancho1982,
author={Sancho, J. M.
and Miguel, M. San
and D{\"u}rr, D.},
title={{Adiabatic elimination for systems of Brownian particles with nonconstant damping coefficients}},
journal={J. Stat. Phys.},
year={1982},
day={01},
volume={28},
number={2},
pages={291-305},
issn={1572-9613},
doi={10.1007/BF01012607},
url={https://doi.org/10.1007/BF01012607}
}

@Article{Hottovy2012,
author={Hottovy, S. and Volpe, G. and Wehr, J.},
title={{Noise-Induced Drift in Stochastic Differential Equations with Arbitrary Friction and Diffusion in the Smoluchowski-Kramers Limit}},
journal={J. Stat. Phys.},
year={2012},
day={01},
volume={146},
number={4},
pages={762-773},
issn={1572-9613},
doi={10.1007/s10955-012-0418-9},
url={https://doi.org/10.1007/s10955-012-0418-9}
}

@article{Brownianinhomenv,
  title = {Brownian motion in inhomogeneous suspensions},
  author = {Yang, M. and Ripoll, M.},
  journal = {Phys. Rev. E},
  volume = {87},
  issue = {6},
  pages = {062110},
  numpages = {7},
  year = {2013},
  publisher = {American Physical Society},
  doi = {10.1103/PhysRevE.87.062110},
  url = {https://link.aps.org/doi/10.1103/PhysRevE.87.062110}
}

@Article{entropyproductionsolvable,
AUTHOR = {Cocconi, L. and Garcia-Millan, R. and Zhen, Z. and Buturca, B. and Pruessner, G.},
TITLE = {{Entropy Production in Exactly Solvable Systems}},
JOURNAL = {Entropy},
VOLUME = {22},
YEAR = {2020},
NUMBER = {11},
ARTICLE-NUMBER = {1252},
URL = {https://www.mdpi.com/1099-4300/22/11/1252},
PubMedID = {33287020},
ISSN = {1099-4300},
DOI = {10.3390/e22111252}
}

@article{hamratchet1,
   title={{Classical and Quantum Hamiltonian Ratchets}},
   volume={87},
   ISSN={1079-7114},
   url={http://dx.doi.org/10.1103/PhysRevLett.87.070601},
   DOI={10.1103/physrevlett.87.070601},
   number={7},
   journal={Phys. Rev. Lett.},
   publisher={American Physical Society (APS)},
   author={Schanz, H. and Otto, M.-F. and Ketzmerick, R. and Dittrich, T.},
   year={2001}}

@article{hamratchet2,
   title={{Intrinsic ratchets: A Hamiltonian approach}},
   volume={98},
   ISSN={2470-0053},
   url={http://dx.doi.org/10.1103/PhysRevE.98.042130},
   DOI={10.1103/physreve.98.042130},
   number={4},
   journal={Phys. Rev. E},
   publisher={American Physical Society (APS)},
   author={Plyukhin, A. V.},
   year={2018}}

@article{dichtratch,
  title = {{Fluctuation driven ratchets: Molecular motors}},
  author = {Astumian, R. D. and Bier, Martin},
  journal = {Phys. Rev. Lett.},
  volume = {72},
  issue = {11},
  pages = {1766--1769},
  numpages = {0},
  year = {1994},
  publisher = {American Physical Society},
  doi = {10.1103/PhysRevLett.72.1766},
  url = {https://link.aps.org/doi/10.1103/PhysRevLett.72.1766}
}

@article{forcedthermalratch,
  title = {Forced thermal ratchets},
  author = {Magnasco, M. O.},
  journal = {Phys. Rev. Lett.},
  volume = {71},
  issue = {10},
  pages = {1477--1481},
  numpages = {0},
  year = {1993},
  publisher = {American Physical Society},
  doi = {10.1103/PhysRevLett.71.1477},
  url = {https://link.aps.org/doi/10.1103/PhysRevLett.71.1477}
}

@ARTICLE{Diffusionrandompotential,
  title    = "Acceleration of diffusion in randomly switching potential with supersymmetry",
  author   = "Dubkov, A. A. and Spagnolo, B.",
  journal  = "Phys Rev E Stat Nonlin Soft Matter Phys",
  volume   =  72,
  number   = "4 Pt 1",
  pages    = "041104",
  month    =  oct,
  year     =  2005,
  address  = "United States",
  language = "en"
}

@article{diffusionrandompotential2,
title = {Nonequilibrium steady-state distributions in randomly switching potentials},
journal = {Phys. A: Stat. Mech. Appl.},
volume = {325},
number = {1},
pages = {26-32},
year = {2003},
note = {Stochastic Systems: From Randomness to Complexity},
issn = {0378-4371},
doi = {https://doi.org/10.1016/S0378-4371(03)00179-1},
url = {https://www.sciencedirect.com/science/article/pii/S0378437103001791},
author = {A. A. Dubkov and P. N. Makhov and B. Spagnolo}
}

\newpage
\begin{center}
{\LARGE \textbf{Appendix}}\\[1ex]
\end{center}
\appendix

\section{Calculation of the reduced dynamics}\label{appendix calculation}
We add details on how to compute the induced mean force, friction coefficient, and noise amplitude.
\subsection{Mean force}
We start by analyzing the Born-Oppenheimer distribution by solving \eqref{bos}. For $\sigma \in \mathbb{Z}_3$, the solution is
\begin{align}
  \rho_x(\sigma) & =  \frac{e^{- \beta \lambda U(x,\sigma)}}{\mathcal{N}_x} g_x(\sigma)\label{bo appendix} \\
  g_x(\sigma) & = \psi_x(\sigma , \sigma + 1) \ \psi_x(\sigma , \sigma + 2) \ e^{\frac{\beta \lambda}{2} \left( U(x, \sigma + 1) + U(x, \sigma + 2) \right)} \nonumber \\
  & \qquad  + \psi_x(\sigma , \sigma + 1) \ \psi_x(\sigma + 1, \sigma + 2) \ e^{\frac{\beta \lambda}{2} \left(U(x, \sigma + 2) + U(x, \sigma) \right)} \ e^{- \beta w} \nonumber \\
  & \qquad + \psi_x(\sigma + 1 , \sigma + 2) \ \psi_x(\sigma , \sigma+2) \ e^{\frac{\beta \lambda}{2} \left( U(x, \sigma + 1) + U(x,\sigma)  \right)} \ e^{\beta w} \label{leading term large w}
\end{align}
with normalization $\cal N_x = \sum_{\sigma \in \mathbb{Z}_{ 3 }}e^{- \beta \lambda U(x,\sigma)} g_x(\sigma) $ and where the sums $\sigma + i$ are taken modulo 3. The equilibrium Boltzmann weight is recovered in the limit $w \to 0$ by noting that $\cal G_x(\sigma) = \lim_{w \to 0} g_x(\sigma)$ is independent of $\sigma$, \textit{i.e.} $\cal G_x(\sigma + 1) = \cal G_x(\sigma)$, such that it can be absorbed into the normalisation factor, leading to
\begin{align}
  \rho_x^{\text{eq}}(\sigma) &= \lim_{w \to 0} \rho_x(\sigma) = \frac{e^{- \beta \lambda U(x,\sigma)}}{Z_x} \label{equilibrium born oppenheimer}
\end{align}
In the small coupling limit $\lambda \ll 1$, \eqref{bo appendix} reduces to
\begin{align}
   \rho_x(\sigma) & = \frac{1}{3} + \frac{\lambda}{3} \Bigg[\frac{h_x(\sigma)}{(1 + 2 \cosh(\beta w))} - \beta U(x,\sigma) - \frac{2}{3} (\Psi_x(0,1) + \Psi_x(1,2) + \Psi_x(2,0)) \Bigg] + O(\lambda^2) \label{rhox weak coupling} \\
   h_x(\sigma) & = \frac{\beta}{2} \left[2 U(x, \sigma) \cosh(\beta w) + U(x, \sigma + 1) \left(1 + e^{\beta w}\right) + U(x, \sigma + 2) \left(1 + e^{- \beta w} \right)\right] \nonumber \\
   & \qquad + \Psi_x(\sigma, \sigma + 1) \left(1 + e^{- \beta w} \right) + \Psi_x(\sigma, \sigma + 2) \left(1 + e^{\beta w} \right) +2 \Psi_x(\sigma + 1, \sigma + 2) \cosh(\beta w) \nonumber 
\end{align}
where we have introduced the average over $\sigma$ as $\langle f(x,\sigma)\rangle_\sigma = \frac{1}{3}\sum_{\sigma \in \mathbb{Z}_3} f(x,\sigma)$.
Hence, under conditions \eqref{mfo}, the relevant component becomes 
\begin{align*}
\rho_x(0) & = \frac{1}{3} - \frac{\lambda}{9(1+2 \cosh (\beta w))} \Bigg[3 \beta (\cosh (\beta w)+1) \tilde{U}(x) \\
& \hspace{6 cm}+ (3 \sinh (\beta w)+\cosh (\beta w)-1) \tilde{\Psi}(x)\Bigg] + O(\lambda^2) \nonumber 
\end{align*}
yielding the mean force in \eqref{bar F weak}
\begin{align*}
  \lambda \bar{F}(x) &= -\lambda \sum_{\sigma \in \mathbb{Z}_3} \tilde{U}'(x) \delta_{\sigma, 0} \ \rho_x(\sigma) = - \lambda \tilde{U}'(x) \ \rho_x(0) \\
  &= -\frac{\lambda N}{3}\tilde{U}'(x) 
  +\lambda^{2} N \frac{(\cosh (\beta w)+1)}{3(2 \cosh (\beta w)+1)}\beta \tilde{U}(x) \ \tilde{U}'(x)
  \\
  & \qquad +\lambda^2 N \frac{(3 \sinh (\beta w)+\cosh (\beta w)-1)}{9(2 \cosh (\beta w)+1)} \tilde{U}'(x) \tilde \Psi (x) 
+O(\lambda^3) \nonumber
\end{align*}
\subsection{Noise amplitude}
Writing out $B(x)$ in \eqref{reduced form D 2} explicitly yields
\begin{align*}
  B(x) & =\lambda^2 N \int_0^{\infty} \id \tau \left[\sum_{\sigma, \sigma_0 \in \mathbb{Z}_n} \frac{\partial U}{\partial x}(x, \sigma) \ \rho_x(\sigma,\tau|\sigma_0,0) \ \frac{\partial U}{\partial x}(x, \sigma_0) \ \rho_x(\sigma_0) - \left(\sum_{\sigma \in \mathbb{Z}_n} \frac{\partial U}{\partial x}(x, \sigma) \ \rho_x(\sigma) \right)^2 \right]
\end{align*}
Here $\rho_x(\sigma,\tau|\sigma_0,0)$ is the transition probability for the $\sigma$-process at fixed $x$ and starting from configuration $\sigma_{0}$ at $t=0$. 

Under weak coupling, since $B(x)$ is already of order $O(\lambda^2)$, we only need the leading order of $\rho_x(\sigma,\tau|\sigma_0,0)$ and $\rho_x(\sigma)$ resulting in
\begin{align*}
  B(x) & =\frac{\lambda^2 N}{3} \sum_{\sigma, \sigma_0 \in \mathbb{Z}_n} \frac{\partial U}{\partial x}(x, \sigma)  \ \frac{\partial U}{\partial x}(x, \sigma_0)  \int_0^{\infty} \id \tau \left(\rho_x^{\lambda = 0}(\sigma,\tau|\sigma_0,0) - \frac{1}{3} \right) + O(\lambda^3)
\end{align*}
In the simplifying case \eqref{only state 1}, this reduces further to
\begin{align*}
  B(x) & =\frac{\lambda^2 N}{3} \tilde{U}'(x)^2 \int_0^{\infty} \id \tau \left(\rho_x^{\lambda = 0}(0,\tau|0,0) - \frac{1}{3} \right) + O(\lambda^3)
\end{align*}
The relevant transition matrix element is given by
\begin{align}
  \rho_x^{\lambda = 0}(0,\tau|0,0) &= \frac{1}{3} \left[1+2 e^{-3 \tau \psi _0 \cosh \left(\frac{\beta w}{2}\right)} \cos \left(\sqrt{3} \tau \psi _0 \sinh \left(\frac{\beta w}{2}\right)\right)\right] \label{00 component transition} \\   
  \int_0^{\infty} \id \tau \left(\rho_x^{\lambda = 0}(0,\tau|0,0) - \frac{1}{3} \right) & = \frac{2 \cosh \left(\frac{\beta w}{2}\right)}{3 \psi_0 \left(2 \cosh (\beta w)+1\right)} \nonumber 
\end{align}
resulting in
\begin{align*}
  B(x) & =\frac{2 \lambda^2 N \cosh \left(\frac{\beta w}{2}\right)}{9 \psi_0 \left(2 \cosh (\beta w)+1\right)} \tilde{U}'(x)^2  + O(\lambda^3)
\end{align*}
in agreement with formula \eqref{B weak coupling result}. 
\subsection{Friction coefficient}
The covariance $\langle \cdot \ ; \ \cdot \rangle^{\text{BO}}_x$ in the formula \eqref{reduced form nu s} for $\nu$ can be rewritten as a single expectation value
\begin{align}
  \nu(x) &= - \lambda N \int_0^{\infty} \id \tau \ \left \langle \frac{\partial U}{\partial x}(x(t), \sigma(\tau)) \cdot  \frac{\partial \log \rho_x}{\partial x}(\sigma) \right \rangle_x^{\text{BO}} \nonumber \\
  & \qquad + \lambda N \int_0^\infty \id \tau \ \left \langle\frac{\partial U}{\partial x}(x(t), \sigma(\tau)) \right\rangle_x^{\text{BO}} \cdot \left \langle \ \frac{\partial \log \rho_x}{\partial x}(\sigma)  \right \rangle_x^{\text{BO}} \nonumber \\
  & = - \lambda N \int_0^{\infty} \id \tau \ \left \langle \frac{\partial U}{\partial x}(x(t), \sigma(\tau)) \cdot  \frac{\partial \log \rho_x}{\partial x}(\sigma) \right \rangle_x^{\text{BO}} \label{reduced form nu 2}
\end{align}
where we have used the normalization of $\rho_x$ to eliminate the second term since
\begin{align*}
  \left \langle \frac{\partial \log \rho_x}{\partial x}(\sigma)  \right \rangle_x^{\text{BO}} & = \sum_{\sigma \in \mathbb{Z}_n}\frac{\partial \rho_x}{\partial x}(\sigma) = 0
\end{align*}
Writing out the last expression in \eqref{reduced form nu 2} yields the form
\begin{align*}
  \nu(x) & = - \lambda N \int_0^{\infty} \id \tau \ \sum_{\sigma, \sigma_0 \in \mathbb{Z}_n} \frac{\partial U}{\partial x}(x(t), \sigma) \ \rho_x(\sigma, t | \sigma_0,0)   \frac{\partial \rho_x}{\partial x}(\sigma_0) 
\end{align*}
Under weak coupling, we have from \eqref{rhox weak coupling} that $ \frac{\partial \rho_x}{\partial x}(\sigma_0) = O(\lambda)$, making $\nu(x)$ of order $O(\lambda^2)$, such that we only need the leading order of $\rho_x(\sigma, \tau| \sigma_0,0)$, \textit{i.e.}
\begin{align*}
  \nu(x) & = - \frac{\lambda^2 N}{3} \int_0^{\infty} \id \tau \sum_{\sigma, \sigma_0 \in \mathbb{Z}_n} \frac{\partial U}{\partial x}(x, \sigma) \ \rho_x^{\lambda = 0}(\sigma, \tau| \sigma_0,0) \ \cdot \\ 
  & \qquad \Bigg[\frac{\partial_x h_x(\sigma_0)}{(1 + 2 \cosh(\beta w))} - \beta \partial_x U(x,\sigma_0) - \frac{2}{3} (\partial_x\Psi_x(0,1) + \partial_x\Psi_x(1,2) + \partial_x\Psi_x(2,0)) \Bigg] + O(\lambda^3) \nonumber 
\end{align*}
In the simplifying case \eqref{only state 1}, this reduces further to \begin{align*}
  \nu(x) & = - \frac{\lambda^2 N}{3 (1 + 2 \cosh(\beta w))} \tilde{U}'(x) \int_0^{\infty} \id \tau \ \cdot  \\
  & \Bigg[\frac{\beta \tilde{U}'(x)}{2} \left(-\rho_x^{\lambda = 0}(0, \tau| 0,0) 2(1+\cosh(\beta w)) + \rho_x^{\lambda = 0}(0, \tau| 1,0) (1 + e^{- \beta w}) + \rho_x^{\lambda = 0}(0, \tau| 2,0) (1 + e^{\beta w})\right) \\
  & \quad + \frac{\tilde{\Psi}'(x)}{3} \Bigg(\rho_x^{\lambda = 0}(0, \tau| 0,0) (1 - \cosh(\beta w) - 3 \sinh(\beta w)) + \rho_x^{\lambda = 0}(0, \tau| 1,0) (1 -\cosh(\beta w) + 3 \sinh(\beta w)) \\
  & \hspace{4 cm} + \rho_x^{\lambda = 0}(0, \tau| 2,0) 2 ( \cosh(\beta w)-1) \Bigg) \Bigg] + O(\lambda^3) \nonumber 
\end{align*}
The necessary elements are \eqref{00 component transition} and
\begin{align*}
  \rho_x^{\lambda = 0}(0,\tau|1,0) &= \frac{1}{3}\left[ 1-e^{-3 \tau \psi _0 \cosh \left(\frac{\beta w}{2}\right)} \left(\sqrt{3} \sin \left(\sqrt{3}
  \tau \psi _0 \sinh \left(\frac{\beta w}{2}\right)\right)+\cos \left(\sqrt{3} \tau \psi _0 \sinh
  \left(\frac{\beta w}{2}\right)\right)\right) \right] \\
  \rho_x^{\lambda = 0}(0,\tau|2,0) &= \frac{1}{3} \left[1+e^{-3 \tau \psi _0 \cosh \left(\frac{\beta w}{2}\right)} \left(\sqrt{3} \sin \left(\sqrt{3} \tau 
  \psi _0 \sinh \left(\frac{\beta w}{2}\right)\right)-\cos \left(\sqrt{3} \tau \psi _0 \sinh \left(\frac{\beta 
  w}{2}\right)\right)\right)\right]
\end{align*}
Performing the integrals yields
\begin{align*}
  &\int_0^\infty \id \tau \left(-\rho_x^{\lambda = 0}(0, \tau| 0,0) 2(1+\cosh(\beta w)) + \rho_x^{\lambda = 0}(0, \tau| 1,0) (1 + e^{- \beta w}) + \rho_x^{\lambda = 0}(0, \tau| 2,0) (1 + e^{\beta w})\right) \\
  & = -\frac{4 \cosh \left(\frac{\beta w}{2}\right) (\cosh (\beta w)+2)}{3 \psi _0 (2 \cosh (\beta w)+1)}\\
   &\int_0^\infty \id \tau \Bigg(\rho_x^{\lambda = 0}(0, \tau| 0,0) (1 - \cosh(\beta w) - 3 \sinh(\beta w)) + \rho_x^{\lambda = 0}(0, \tau| 1,0) (1 -\cosh(\beta w) + 3 \sinh(\beta w)) \\
   & \hspace{4 cm}+ \rho_x^{\lambda = 0}(0, \tau| 2,0) 2 ( \cosh(\beta w)-1) \Bigg) \\
  & = -\frac{e^{-\frac{\beta w}{2}} \left(e^{2\beta w}+ e^{ \beta w}-2\right)}{\psi _0 (2 \cosh (\beta w)+1)}
\end{align*}
As a consequence,
\begin{align*}
  \nu(x) & = \frac{\lambda^2 N}{9 \psi_0 (1 + 2 \cosh(\beta w))^2} \Bigg[2\beta \cosh \left(\frac{\beta w}{2}\right) (\cosh (\beta w)+2)\tilde{U}'(x)^2 \\
  & \hspace{6 cm} + e^{-\frac{\beta w}{2}} \left(e^{2\beta w}+ e^{ \beta w}-2\right)\tilde{U}'(x) \tilde{\Psi}'(x)\Bigg] + O(\lambda^3)
\end{align*}
which is \eqref{frictionCoeff-WCoup}. For the specific case \eqref{only state 1}, that reduces to
 \begin{equation*}
     \nu(x) = \frac{A + B \cos(\varphi)}{2} + R \cos\left(\frac{4 \pi x}{L} - \delta \right)
     \end{equation*}
     with 
     \begin{align*}
     R &= \frac{1}{2} \sqrt{A^2+B^2 + 2 A B \cos(\varphi)}, \qquad \tan(\delta) = -\frac{B \sin(\varphi)}{A + B \cos(\varphi)} \\
     A&=\frac{8 \pi^2\lambda^2}{9(2 \cosh (\beta w)+1)^2} \cosh \left(\frac{\beta w}{2}\right) (\cosh (\beta w)+2) \frac{\beta U_0^2}{\psi_0 L^2}\,, \\
     B & = \frac{4 \pi^2\lambda^2}{9 (2 \cosh (\beta w)+1)^2}e^{-\frac{\beta w}{2}} \left(e^{2\beta w}+ e^{ \beta w}-2\right) \frac{U_0 \Psi_0}{\psi_0 L^2}\,.
   \end{align*}
from which the conditions \eqref{eq:negfric_negative_w}--\eqref{eq:negfric_positive_w} of positive or negative friction can be derived.
\subsection{Large driving limit}
In the large driving limit $w \to \infty$, only the term \eqref{leading term large w} in $\rho_x$ survives such that
\begin{align*}
  \lim_{w \to \infty }\rho_x(\sigma) & \propto e^{\frac{\beta \lambda}{2} \left( U(x, \sigma + 1) - U(x,\sigma)  \right)} \lim_{w \to \infty } \psi_x(\sigma + 1 , \sigma + 2 ; w ) \ \psi_x(\sigma , \sigma+2 ; w ) \ 
\end{align*}
 Using that the combination $\psi_x(\sigma, \sigma + 1; w) \ \psi_x(\sigma + 1 , \sigma + 2; w ) \ \psi_x(\sigma , \sigma+2 ; w)$ is independent of $\sigma$ and can be absorbed into the proportionality constant, this can also be written as in \eqref{boslim}. A similar analysis shows that for $w \to - \infty$
\begin{align*}
  \lim_{w \to - \infty } \rho_x(\sigma) & \propto  e^{\frac{\beta \lambda}{2} \left( U(x, \sigma + 2) - U(x,\sigma)  \right)} \lim_{w \to \infty }\frac{1}{\psi_x(\sigma + 2, \sigma )}
\end{align*}
For the friction and noise, using the property $\left \langle X \ ; \ Y \right \rangle_x^{\text{BO}} = \left \langle \left( X - \left \langle X \right \rangle_x^{\text{BO}}\right)  Y \right \rangle_x^{\text{BO}}$, the expressions in \eqref{reduced form nu s}--\eqref{reduced form D 2} can be rewritten in the form
\begin{align*}
 \left \langle g_x(\sigma; w) \int_0^\infty \id \tau\left[e^{\tau \cal L_\sigma}f_x(\sigma)- \langle f_x(\sigma)\rangle^\text{BO}_x\right] \right \rangle_x^{\text{BO}}
\end{align*}
with $f_x(\sigma) = \partial_x U(x,\sigma)$ while $g_x(\sigma; w) = \partial_xU(x,\sigma)$ for the nosie and $g_x(\sigma; w) = \partial_x \log \rho_x(\sigma; w)$. We also introduced the backward generator $\cal L_\sigma$ of the Markov jump process defined as the transpose of the generator in \eqref{bos}, and acting on the $\sigma$-part of functions $f_x(\sigma)$. Since in both cases the limit $G_x(\sigma) = \lim_{w \to \pm \infty} g_x(\sigma;w)$ exists, we have that the large $|w|$-limit is dominated by the behavior of
\begin{align}
\lim_{w \to \pm \infty} \int_0^\infty \id \tau \left[e^{\tau \cal L_\sigma}f- \langle f\rangle^\text{BO}_x\right]= -\lim_{w \to \pm \infty} \cal L_\sigma^{-1} [f - \langle f\rangle^\text{BO}_x] \label{inverse L appendix}  
\end{align}
This last expression is typically dominated by the first nonzero eigenvalue $\lambda_1$ of $\mathcal L_\sigma$, which then decides whether the friction and the noise vanish or not for very large chemical bias, at least when there is a well-defined limit $\lim_{w \to \infty} \langle f\rangle^\text{BO}_x $
of the Born-Oppenheimer density \eqref{bos} in \eqref{boslim}. As a specific illustration of \eqref{inverse L appendix}, let us for simplicity consider the case \eqref{only state 1} at positive driving $w \to \infty$ and rewrite the generator in the matrix form $\mathcal L_\sigma(w) = e^{\frac{\beta w}{2}} \psi_0 Q_\sigma(w)$ where
\begin{align*}
  Q_\sigma(w) & = \left(
\begin{array}{ccc}
 -\left(\frac{\psi(0,1)}{\psi_0} e^{\beta w}+1\right) e^{\frac{1}{2} \beta \lambda \tilde{U}(x)-\beta w} & \frac{\psi(0,1)}{\psi_0} e^{\frac{1}{2} \beta \lambda \tilde{U}(x)} & ^{\frac{1}{2} \beta \lambda \tilde{U}(x)-\beta w}\\
 \frac{\psi(0,1)}{\psi_0} 
  e^{-\frac{1}{2} \beta (\lambda \tilde{U}(x)+2 w)} & -\frac{\psi(0,1)}{\psi_0} e^{-\frac{1}{2} \beta (\lambda \tilde{U}(x)+2 w)}-1 &
  1 \\
 e^{-\frac{1}{2} \beta \lambda \tilde{U}(x)}  & e^{-\beta w} & -e^{-\frac{1}{2} \beta \lambda \tilde{U}(x)}-e^{-\beta w} \\
\end{array}
\right)
\end{align*}
The important point is that we do not need the matrix exponential directly, but only its integral, for which we can freely rescale time as $s = e^{\frac{\beta w}{2}} \psi_0 \tau$, yielding
\begin{align*}
   \ \int_0^\infty \id \tau \left( e^{\tau \mathcal L_\sigma(w)} f_x(\sigma;w) - \left \langle f_x(\sigma; w) \right \rangle_x^{\text{BO}} \right) = \frac{e^{- \frac{\beta w}{2}}}{\psi_0} \int_0^\infty \id s  \left( e^{Q_\sigma(w) s} f_x(\sigma;w) - \left \langle f_x(\sigma; w) \right \rangle_x^{\text{BO}} \right) 
\end{align*}
since $e^{\frac{\beta w}{2}} \psi_0 > 0$. As $\mathcal L_\sigma$ has non-positive eigenvalues, the same holds for $Q_\sigma$ such that the last integral converges. Next, assuming the limits $\lim_{w \to \infty} \psi_x(\sigma, \sigma' ; w)$ are finite, we have 
\begin{align*}
   &Q_\sigma^{\infty} = \lim_{w \to \infty} Q_\sigma(w) = \left(
\begin{array}{ccc}
 -\frac{\lim_{w \to \infty}\psi_x(0,1)}{\psi_0} e^{\frac{1}{2} \beta \lambda \tilde{U}(x)} & \frac{\lim_{w \to \infty}\psi_x(0,1)}{\psi_0} e^{\frac{1}{2} \beta \lambda \tilde{U}(x)} & 0\\
 0 & -1 & 1 \\
 e^{-\frac{1}{2} \beta \lambda \tilde{U}(x)} & 0 & -e^{-\frac{1}{2} \beta \lambda \tilde{U}(x)} \\
\end{array}
\right)\\
&\lim_{w \to \infty }e^{Q_\sigma(w) s} = e^{Q_\sigma^\infty s}
\end{align*}
where the last equality follows from the continuity of the matrix exponential function. Consequently, assuming $f_x^\infty(\sigma) = \lim_{w \to \infty} f_x(\sigma; w)$ exists, it follows from $ \lim_{w \to \infty} e^{- \beta w/2} = 0$ that
\begin{align*}
   & \lim_{w \to \infty} \int_0^\infty \id \tau \left( e^{\tau \mathcal L_\sigma(w)} f_x(\sigma;w) - \left \langle f_x(\sigma; w) \right \rangle_x^{\text{BO}} \right) \\
   & = \lim_{w \to \infty}\frac{e^{- \frac{\beta w}{2}}}{\psi_0} \int_0^\infty \id s \left( e^{Q_\sigma^\infty s}f_x^\infty(\sigma) - \left \langle f_x^\infty(\sigma) \right \rangle_x^{\text{BO}, \infty}\right) = 0
\end{align*}
such that the noise and friction vanish for $w \to \infty$.

\section{Numerical scheme}\label{appendix numerics}
We integrate the equations of motion for the case \eqref{only state 1}, \eqref{phase difference} using a stochastic algorithm that combines a deterministic mechanical update for the probe with a continuous-time Markov chain jump process for the bath variables. \\

The mechanical degrees of freedom, particle position $x$ and velocity $v$, are evolved using the semi-implicit Euler method. At each time step of size 
$\Delta t$, the acceleration is first evaluated at the current state $(x_i, \vec{\sigma}_i)$,
\begin{equation}\label{ai}
  a_i = \frac{2 \pi}{m L}\left[-V_0 \sin\left(\frac{2 \pi}{L} x_i \right) + N_0^i\, U_0 \sin\left(\frac{2 \pi}{L} x_i\right)\right]
\end{equation}
where $N_0^i = \sum_{j=1}^{N} \delta_{\sigma_{j,i}, 0}$ is the number of 
internal units in state $0$ at timestep $i$. The velocity and position are then updated as
\begin{align*}
  v_{i+1} &= v_i + a_i \,\Delta t, \qquad x_{i+1} = \bigl(x_i + v_{i+1}\,\Delta t\bigr) \bmod L,
\end{align*}
where $\bmod \ L$ enforces periodic boundary conditions on the domain circle $[0, L)$. For the case on the real line from section \eqref{section real line}, one instead uses the potentials \eqref{potential real line} to calculate the acceleration, and no modulo operation is applied. \\

Next, after the mechanical update, the internal state vector $\vec{\sigma}_i = (\sigma_{1,i}, \ldots, \sigma_{N,i}) \in 
\{0,1,2\}^N$ evolves as $N$ independent continuous-time Markov chains. Defining the escape rate
\begin{equation*}
  \Lambda_x(\sigma) = \sum_{\sigma' \neq \sigma} k_x(\sigma, \sigma')
\end{equation*}
we have that over a time interval $\Delta t$, the probability that jumper $\sigma_j$ (currently in state $\sigma_{j,i}$) undergoes at least one transition, is approximated by 
\begin{equation*}
  p_{j,i} = 1 - e^{-\Lambda_{x}(\sigma_{j,i})\,\Delta t}
\end{equation*}
This is exact for a two-state system and is a standard approximation for multi-state systems under the assumption that $\Delta t$ is small enough that at most one jump occurs per unit per step. Then, every jumper $\sigma_j$ is assigned a uniform random variate $u_j \sim \mathcal{U}(0,1)$ such that $\sigma_j$ jumps if $u_j < p_{j,i}$ and remains the same otherwise. Whenever a jump occurs, the destination state is selected by drawing $r_j \sim \mathcal{U}(0,\, \Lambda_{x}(\sigma_{j,i}))$ and 
comparing against the rate $k_x(0,1)$. For example, for a jumper $\sigma_j$ in state $0$, it transitions to $1$ if $r_j < k_x(0,1)$, and to $2$ otherwise. \\

In conclusion, the full update at each step $i \to i+1$ proceeds as follows:
\begin{enumerate}
  \item Compute the acceleration $a_i$ in \eqref{ai} from the current state $(x_i, \vec{\sigma}_i)$.
  \item Update the velocity: $v_{i+1} = v_i + a_i\,\Delta t$.
  \item Update the position: $x_{i+1} = (x_i + v_{i+1}\,\Delta t) \bmod L$.
  \item Evaluate the transition rates $k_{x_{i+1}}(\sigma, \sigma')$ at the new position.
  \item For each jumper $\sigma_j$, independently sample whether a jump occurs and, if so, the destination state.
  \item Store $(x_{i+1},\, v_{i+1},\, \vec{\sigma}_{i+1})$.
\end{enumerate}
\end{document}